\documentclass[prx,twocolumn,english,
,superscriptaddress,
floatfix,longbibliography]{revtex4-2}

\pdfoutput=1
\usepackage[utf8]{inputenc}
\usepackage[T1]{fontenc}
\usepackage{braket,xcolor}
\usepackage{graphicx,enumerate,verbatim,bbold}
\usepackage{amsmath,amssymb,amsthm,float,mathrsfs}
\usepackage{dsfont}
\usepackage[normalem]{ulem}
\usepackage{lmodern}
\usepackage{hyperref}
\hypersetup{colorlinks,citecolor=blue,linkcolor=blue,urlcolor=blue}
\usepackage{braket}
\usepackage{multirow}
\usepackage{bbm}
\usepackage{physics}
\usepackage{bm}
\usepackage{tikz}
\usetikzlibrary{patterns}
\usetikzlibrary{fadings}
\usetikzlibrary{quantikz2}
\usepackage[normalem]{ulem}
\usepackage[caption=false]{subfig}
\usepackage{dcolumn}

\newcommand{\bqa}{\begin{eqnarray}}
\newcommand{\eqa}{\end{eqnarray}}

\newcommand{\be}{\begin{equation}}
\newcommand{\ee}{\end{equation}}

\newcommand{\f}{\text{f}}

\newcommand{\Htarget}{K_{\mathbf{T}}}
\newcommand{\Hinit}{K_0}
\newcommand{\rhotarget}{\varrho_{\mathbf{T}}^\beta}
\newcommand{\tf}{t_\f}

\newcommand{\del}[1]{\hil{{\bf XXX}}}

\newcommand{\paul}[1]{{\color[rgb]{0.0, 0.4, 0.6}{#1}}}
\newcommand{\flo}[1]{{\color[rgb]{0,0,0.6}{#1}}}
\newcommand{\alex}[1]{{\color[rgb]{0,0.6,0}{#1}}}
\newcommand{\hil}[1]{{\color[rgb]{0.6,0,0}{#1}}}
\newcommand{\um}{\mathbbm{1}}
\usepackage{subfig}

\definecolor{nqdcolor}{rgb}{0.5586, 0.0586, 0.4219}
\newcommand*{\nqdcolor}{\color{nqdcolor}}
\usepackage{comment}
\newcommand{\mb}[1]{{\nqdcolor{#1}}}

\newcommand{\Imperial}{Blackett Laboratory, Imperial College London, SW7 2AZ, United Kingdom}
\newcommand{\PKS}{Max Planck Institute for the Physics of Complex Systems, MPI-PKS, N\"othnitzerstr.~38, 01187 Dresden, Germany}

\newcommand{\Qolumbus}{Qolumbus SAS, 78 Boulevard de la République, 92100 Boulogne-Billancourt, France}

\begin{document}
\title{
Fast thermal state preparation beyond native interactions
}


\author{Alexander van Lomwel}\affiliation{\Imperial}
\author{Paul M.~Schindler}\affiliation{\PKS}
\author{Modesto Orozco-Ruiz}\affiliation{\Imperial}
\author{Marin Bukov}\affiliation{\PKS}
\author{Nguyen H. Le}\affiliation{\Imperial}\affiliation{\Qolumbus}
\author{Florian Mintert}\affiliation{\Imperial}


\begin{abstract}

While questions on quantum simulation of ground state physics are mostly focussed on the realization of effective interactions,
most work on quantum simulation of thermal physics explores the realization of dynamics towards a thermal mixed state under native interactions.
Many open questions that could be answered with quantum simulations, however, involve thermal states with respect to synthetic interactions.\\
We present a framework based solely on unitary dynamics to design quantum simulations for thermal states with respect to Hamiltonians that include non-native interactions, suitable for both present-day digital and analogue devices. By classical means, our method finds the control sequence to reach a target thermal state for system sizes well out of reach of state-vector or density-matrix control methods, even though quantum hardware is required to explicitly simulate the thermal state dynamics.
With the illustrative example of the cluster Ising model that includes non-native three-body interactions, we find that required experimental resources, such as the total evolution time, are independent of temperature and criticality. 
\end{abstract} 
\maketitle


\begin{figure*}
  \centering
  \includegraphics[width=0.9\linewidth]{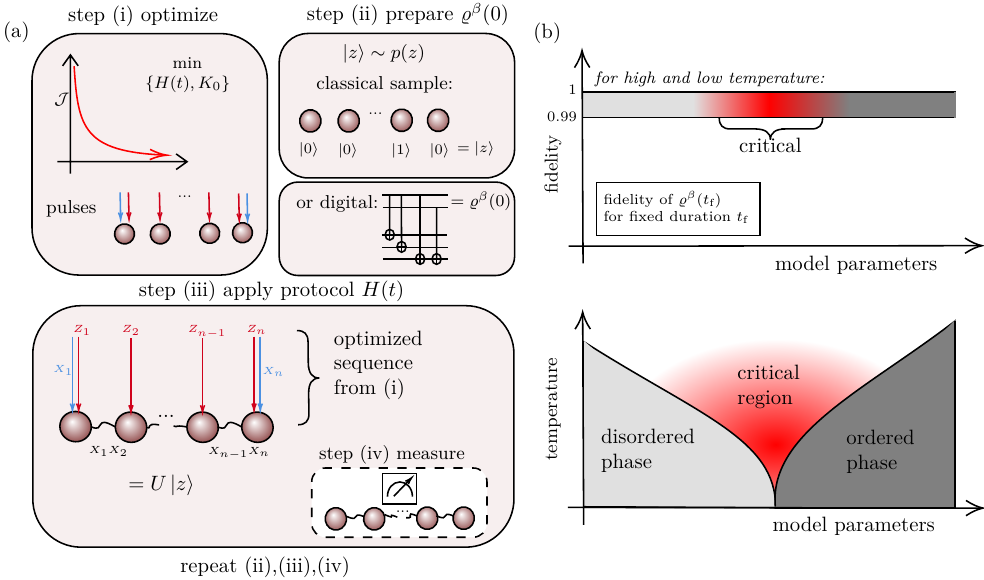}
  \caption{The left panel (a) summarises the target thermal state $\varrho_\mathbf{T}^\beta$ preparation as a four-step process. (i) An optimization, with respect to a figure of merit $\mathcal{J}$, to determine the specific time-dependencies of the system Hamiltonian, $H(t)$, and the initial condition $\Hinit$ (Eq.~\eqref{eq:optimization}). (ii) The preparation of the thermal state of the initial condition $\Hinit$, $\varrho^\beta(0)$, either by sampling computation basis states as a chain of excited/unexcited spins, $\ket{z}$, from the corresponding Boltzmann distribution, $p(z)$, (suited to analogue devices), or by a gate sequence (suited to digital devices). (iii) The optimized control sequence, realized by dynamics induced by $H(t)$ found in (i), is applied to the spin-chain, or implemented via gates in the case of digital devices. (iv) An observable of interest is measured. Steps (ii),(iii),(iv) are repeated as many times as are needed to accurately capture the statistics. \\
The right panel (b) shows a sketch of what can be achieved using the present method. The results in Sec.~\ref{sec:results} demonstrate that one can prepare the unitarily evolved thermal state $\varrho^\beta(\tf)\approx \varrho^\beta_\mathbf{T}$ for a broad range of temperatures, even in critical regions of a phase diagram. Strikingly, for a given number of spins, the required evolution time $\tf$ (to reach a state within some accuracy of the target) appears to be unaffected by criticality. As such, the results show that one can prepare high-fidelity thermal states (typically $\sim0.99$ in the worst case) even in critical regimes and low temperature.}
\label{fig:scheme}
\end{figure*}


\section{Introduction}

How to prepare finite-temperature quantum states on quantum simulation devices is a foundational question with wide ranging implications for the applicability of quantum simulators.
It gives controlled access to equilibrium thermodynamics directly from microscopic Hamiltonians, and indirectly to non-equilibrium thermodynamics via Jarzynski's equality~\cite{jarzynski1997nonequilibrium}, provides a clean arena to test thermalization via the Eigenstate Thermalization Hypothesis \cite{d2016quantum}, and enables direct estimation of free energies and observables that organize phase diagrams~\cite{sagastizabal2021variational, chen2023quantum}.
Practically, accurate thermal state preparation on quantum simulators enables us to probe regimes that remain challenging for classical methods ({\it e.g.}, where quantum Monte Carlo faces a sign problem~\cite{loh1990sign, troyer2005computational}).

Common approaches to preparing thermal states as steady-states of dissipative evolution include using effective Lindblad  dynamics~\cite{sweke2014simulation,dive2015quantum, barreiro2011open,chen2023quantum,chen2023efficient,rall2023thermal,guo2025designing,hahn2025provably,hahn2025efficient}, or coupling the system weakly to controlled bath degrees of freedom~\cite{hagan2025thermodynamic, langbehn2025universal, lloyd2025quantum, scandi2025thermalization, shtanko2021preparing}, see \cite{ding2025efficient} for an overview.
While these protocols have become increasingly mature, many open questions, such as the required protocol durations to reach a desired fidelity, remain.
Yet, there is growing evidence that thermal states at low temperature or at criticality require particularly long times to converge to the steady-state, whereas convergence is often much faster at sufficiently high temperature, or far from quantum phase transitions (non-criticality)~\cite{temme2011quantum,kastoryano2013quantum}.

These preparation protocols focus predominantly on thermal state preparation for early fault-tolerant quantum computers; this requires a complete gate set and ancillary degrees of freedom, which makes them inappropriate for analog quantum simulators.
Existing protocols for these devices focus on thermal state preparation with respect to native interactions: this sharply contrasts with analog quantum simulators primarily targetting physics of Hamiltonians with non-native interactions--such as three-body interactions.
Many of the actual physics problems that we aim to address in quantum simulation require the ability to both engineer synthetic interactions and to ensure evolution to the desired thermal state in the presence of these interactions~\cite{jordan2008perturbative, georgescu2014quantum, eckardt2017colloquium}.

In this work, we present a thermal state preparation framework that is compatible with synthetic interactions and avoids the need for engineered dissipative dynamics---making it uniquely suited to both digital {\it and} analogue quantum simulation devices.
This is achieved in terms of unitary dynamics, such that an initial thermal state, that can be prepared with reasonable effort, evolves towards a thermal state with respect to a target Hamiltonian.
Importantly, this Hamiltonian may well include interactions that are not native to the system used for implementation.
The corresponding unitary dynamics will be found as the solution of an optimal control problem that is independent of the temperature of the thermal state.
As such, the present approach applies to the full range of temperatures.

We exemplify our protocol on the paradigmatic cluster-Ising model: an iconic model with a three-body interaction that does not exist as a natural interaction in any quantum device.
Cluster states, which arise as the ground states of the cluster–Ising model, appear at finite temperature in contexts ranging from measurement-based quantum computing to symmetry-protected topological (SPT) order, and their thermalization properties remain an active area of research~\cite{smacchia2011statistical, raussendorf2003measurement}.
The SPT phase and the Ising-ordered phase are separated by a topological quantum phase transition~\cite{ding2019phase}.
This model is thus ideally suited to demonstrate how the present method provides the theoretical foundation for thermal state preparation both at criticality and at low temperatures, {\it i.e.}, under conditions that proved to be challenging in earlier approaches.
Notably, we observe that the duration of system dynamics required for thermal state preparation is also largely independent of aspects like criticality, and is mostly limited by the fundamental quantum speed limit set by the intrinsic system interaction. This stands in sharp contrast to dissipative evolution protocols, for which the required evolution time generally increases substantially in the vicinity of a critical point.

Our protocol provides a reliable thermal state preparation protocol ideally suited to the constraints of present-day quantum simulators---relying only on unitary dynamics generated by native interactions.

\section{A bird's eye view on thermal state preparation}
\label{sec:bird}

Let us first give an overview of our thermal state preparation protocol.
Our goal is to prepare the thermal density matrix
\begin{equation}
\label{eq:targetstate}
\varrho_{\mathbf{T}}^\beta=
    \frac{\exp(-\beta \Htarget)}{\tr \exp(-\beta \Htarget)}\ ,
\end{equation}
for the target Hamiltonian $\Htarget$ at inverse temperature $\beta$, evolved from an experimentally realizable initial state (cf.~Sec.~\ref{sec:initialstate}). Respecting present-day quantum simulation capabilities, we focus on \textit{unitary-only} protocols generated by a system Hamiltonian $H(t)$ constrained to experimentally available controls---below we consider constant nearest-neighbour interactions and tunable time-dependent single-qubit terms.
In particular, the target Hamiltonian $\Htarget$ may not lie within the set of available system Hamiltonians.

In addition, we consider starting from an easy-to-initialize thermal state $\rho^\beta(0) \propto \exp(-\beta \Hinit)$---with initial parent Hamiltonian $\Hinit$.
Note that the initial state is prepared at the same inverse temperature $\beta$ as the target state, since unitary evolution preserves thermal occupation,
\begin{equation}
\label{eq:thermalstatemap}
    \varrho^\beta(t)=U(t)\ \varrho^\beta(0)\  U^\dagger(t) \propto e^{- \beta K(t)} \,,
\end{equation}
{\it i.e}., only the parent Hamiltonian $K(t)$ changes under the evolution $U(t) = \mathcal{T}\mathrm{exp}\left(-i\int^t_0\mathrm{d}s H(s)\right)$, with $K(0)=\Hinit$ and
\begin{equation}
    K(t)=U(t)\ \Hinit \ U^\dagger(t)\,   .
\label{eq:Isol}
\end{equation}
Thus, to prepare the target state $\rhotarget=\varrho^\beta(\tf)$ at final time $\tf$ using unitary evolution, we have to find an experimental protocol $H(t)$ that propagates a suitable initial condition $\Hinit$ to the desired target Hamiltonian $K(\tf)=\Htarget$, following the von Neumann equation, cf. \footnote{Eq.~\eqref{eq:Isol} might look deceptively similar to the evolution $A(t)=U(t)^\dagger A(0)U(t)$ of an observable $A$ in the Heisenberg picture, all of the following discussion applies to the Schrödinger picture with time-independent observables.}, 
\begin{equation}
\dot K(t)=-i[H(t),K(t)].
\label{eq:vN}
\end{equation}

The numerical integration of the von Neumann equation (Eq.~\eqref{eq:vN}) is typically prohibited by the exponential growth of a system's Hilbert space with the number of interacting degrees of freedom.

A prominent exception to this rule is given when the system Hamiltonian is comprised exclusively of elements of a Lie algebra $\mathfrak{b}$, {\it i.e.} a set of operators $\{b_j\}$ with closed commutation relations
\be
    \left[b_j ,b_k\right]=-i\sum _{l=1}^d \lambda _{lj}^k b_l\ ,
    \label{eq:Lie}
\ee
in terms of scalar structure constants $\lambda _{lj}^k$, 
where the number of elements ({\it i.e.} $d$) grows sub-exponentially with the number of interacting degrees of freedom, and, is thus smaller than the dimension of the underlying Hilbert space.
If the initial condition $K(0)$ can be expanded in terms of elements of the Lie algebra, the solution of Eq.~\eqref{eq:vN}, {\it i.e.}
$K(t)$ in Eq.~\eqref{eq:Isol}, can be constructed by integrating $d$ coupled differential equations,
even though construction of the propagator $U(t)$ or the explicit multiplication of $K(0)$ and $U(t)$ in Eq.~\eqref{eq:Isol} requires an exponentially large effort~\cite{orozco2024quantum}.

Even if the dynamics $U(t)K(0)U^\dagger(t)$ takes place in a low-dimensional space spanned by the elements of the Lie algebra, the dynamics $U(t)\varrho^{\beta}(0) U^\dagger(t)$ is typically not restricted to this low-dimensional space, because the initial thermal state $\varrho^{\beta}(0)$ includes powers of $\Hinit$, which are not necessarily elements of the Lie algebra.
The ability to 
integrate Eq.~\eqref{eq:vN} efficiently with the initial condition $K(0)=\Hinit$, thus does not typically imply that Eq.~\eqref{eq:vN} can be integrated efficiently with an initial condition $K(0)\propto\exp(-\beta \Hinit)$.
Even though Eq.~\eqref{eq:vN} with the initial condition $K(0)=\Hinit$ can be used to find a time-dependent Hamiltonian $H(t)$ that realises the desired state preparation, 
the time evolution with a general initial condition -- including the initial thermal states $\varrho^\beta(0)$ -- can not be simulated classically for large enough systems, but it needs to be implemented on an actual quantum device.

While the restriction to a small Lie algebra enables the explicit design of system Hamiltonians with optimized time dependence for system sizes that are far out of reach for descriptions in terms of state vectors or density matrices, it poses a fundamental difficulty for optimal control: any pair of isospectral Hermitian operators are related by a general unitary operator; however, the propagators induced by a Hamiltonian comprised of elements of a small Lie algebra is necessarily restricted to the corresponding Lie group.
It is thus crucial to choose the initial operator $K(0)=\Hinit$ in Eq.~\eqref{eq:Isol} such that the target Hamiltonian $\Htarget$ can be reached under the restricted system dynamics.

To this end, it is useful to consider a Cartan decomposition~$\mathfrak{b}=\mathfrak{k}\oplus\mathfrak{m}$ \cite{cartan1894structure,khaneja2000cartan} of the Lie algebra $\mathfrak{b}$
 into two groups of operators $\mathfrak{k}$ and $\mathfrak{m}$, such that 
 \begin{equation}
    \left[\mathfrak{k}, \mathfrak{k}\right]\subset \mathfrak{k},\qquad \left[\mathfrak{m}, \mathfrak{m}\right]\subset \mathfrak{k},\qquad \left[\mathfrak{k}, \mathfrak{m}\right]\subset \mathfrak{m}\ ,
    \label{eq:cartan}
\end{equation}
where $\left[\mathfrak{a}, \mathfrak{b}\right]\subset \mathfrak{c}$ is a short-hand notation stating that $[a,b]\in\mathfrak{c}$ for any $a\in\mathfrak{a}$ and $b\in\mathfrak{b}$.

Within $\mathfrak{m}$, there is a subset $\mathfrak{h}$ that is maximal Abelian, {\it i.e.} all elements of $\mathfrak{h}$ commute with each other, and there is no further element of $\mathfrak{m}$ that commutes with all elements of $\mathfrak{h}$.
There is an existence theorem~\cite{kokcu2022fixed} that states that for any element $m\in\mathfrak{m}$ there is an element $h\in\mathfrak{h}$
and an element $k\in\mathfrak{k}$ such that
\begin{equation}
    m=\exp(ik)h\exp(-ik)\ .
    \label{eq:reach}
\end{equation}
With an initial condition $\Hinit$ defined by $h$ and a target $\Htarget$ defined by $m$ (or vice versa), this theorem provides a guarantee of reachability. As such, there is a guaranteed solution to Eq.~\eqref{eq:Isol} with $K(0)=\Hinit$ and $K(t_f)=\Htarget$ at a final time $t_f$, with a unitary induced by a system Hamiltonian $H(t)$ that generates the Lie algebra $\mathfrak{b}$.

Crucially, the initial condition $\Hinit=h$ is a sum of mutually commuting operators, which is of great help for the preparation of the initial state $\varrho^\beta(0)$. The existence theorem on its own, however, does not specify the initial condition $\Hinit$ beyond the restriction to the set $\mathfrak{h}$ of operators.
The initial condition from which the control target $\Htarget$ is reachable, thus needs to be  determined as part of the optimal control problem, that can be formalised as
\begin{equation}
\begin{aligned}
\min_{H(t),\, \Hinit \in \mathfrak{h}} \quad
& \mathcal{J}\!\big(K(\tf),\, \Htarget\big) \\
\text{s.t.}\quad
& \dot{K}(t) = -\,i\,[H(t),\,K(t)] \ , \\
& K(0) = \Hinit 
\end{aligned}
\label{eq:optimization}
\end{equation}
with an objective function $\mathcal{J}$ that is minimal if and only if the objective is reached, and
with the time-dependent system Hamiltonian $H(t)$ restricted to practically realizable interactions and driving protocols.

The solution of this optimization problem yields the desired time-dependent system Hamiltonian $H(t)$ and constant parent Hamiltonian $\Hinit$.
The experimental implementation of the dynamics following Eq.~\eqref{eq:thermalstatemap} that realizes the desired thermal state, then only requires the ability to prepare the initial state $\varrho^\beta(0)$.

The complete approach can be summarized in the four steps depicted in Fig.~\ref{fig:scheme}a:
\begin{itemize}
\item[(i)] an optimization that yields an initial condition $\Hinit$ and a control sequence that reaches $\Htarget$; 
\item[(ii)] preparation of the initial thermal state $\varrho^\beta(0)\propto\exp(-\beta \Hinit)$;
\item[(iii)] application of the optimized control sequence;
\item[(iv)]measurement of an observable of interest.
\end{itemize}
Sec.~\ref{sec:targetthermalstateprep} discusses this general approach more explicitly with the example of the cluster Ising model.

\section{Thermal state preparation of the Cluster Ising model}
\label{sec:targetthermalstateprep}

The following discussion will exemplify the general approach sketched in Sec.~\ref{sec:bird} for the specific case of the cluster Ising model based on a system Hamiltonian restricted to an extended Ising model.
The system Hamiltonian and control target are defined in more detail in Sec.~\ref{sec:Ising}.
The Lie algebra generated by the individual terms in the system Hamiltonian and its Cartan decomposition, as required for reachability and determination of the initial condition $\Hinit$, is discussed in Sec.~\ref{sec:algebra}.
Explicit solutions of optimal control problems are discussed in Sec.~\ref{sec:results}, and protocols for the preparation of the initial states $\varrho^\beta(0)$ determined by the initial condition are given in Sec.~\ref{sec:initialstate}.

\subsection{System Hamiltonian and cluster Ising model}
\label{sec:Ising}

An explicit implementation of the framework sketched in Sec.~\ref{sec:bird} requires definition of a system Hamiltonian $H(t)$ and a target Hamiltonian $\Htarget$ defining the thermal states to be prepared.

\subsubsection{Cluster Ising model}

\begin{figure}[t]
\centering
\includegraphics[width=0.8\linewidth]{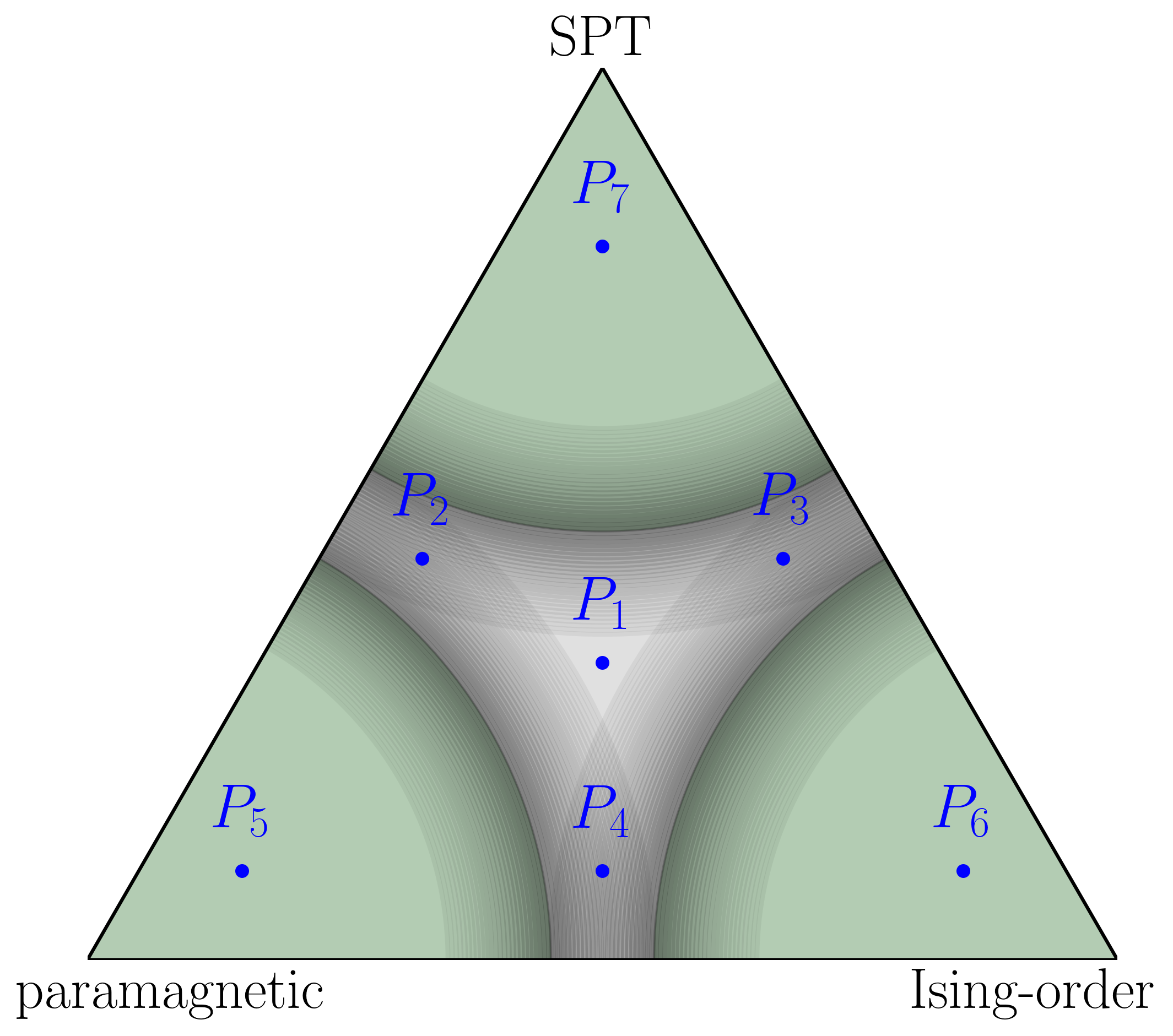}
\caption{
Schematic representation of the phase diagram of the cluster Ising model (Eqs.~\eqref{eq:targetinvariant}, \eqref{eq:scale}),
spanned by the vectors 
$\vec \lambda=(\bm{\lambda},0,0)$ (left bottom),
$\vec \lambda=(0,\bm{\lambda},0)$ (right bottom) and
$\vec \lambda=(0,0,\bm{\lambda})$ (center top)
that define the limiting cases of paramagnetic, Ising-ordered and symmetry-protected topological (SPT) phase.\\
The model exhibits non-critical behaviour with one dominant term in the Hamiltonian, {\it i.e.} domains close to the corners of the triangles depicted in green.
    If two or all three terms in the Hamiltonian are of comparable magnitude (as indicated in grey), the system shows critical behaviour.
    In the asymptotic limit $n\to\infty$, the critical and non-critical regions are separated by a sharp phase transition; for finite system size the crossover between critical and non-critical behaviour takes place over a finite interval in the parameter space, as indicated by the shaded domains.\\
    Seven selected points in the parameter space for which optimal control solutions are discussed in Sec.~\ref{sec:results} are indicated by blue points $P_1$ to $P_7$, with $P_1$ to $P_4$ in critical and $P_5$ to $P_7$ in non-critical regimes.
    }
\label{fig:labelledtriangle}
\end{figure}

To demonstrate the method's ability to reliably prepare thermal states in critical regimes of Hamiltonians including non-native interactions, the target Hamiltonian considered in the following discussion is the cluster Ising Hamiltonian 
\begin{align}
\Htarget = &  \lambda_1 \sum_{j=1}^n Z_j + \lambda_2 \sum_{j=1}^{n-1} X_j X_{j+1}\label{eq:targetinvariant}  \\
    & - \lambda_3\Bigl(Z_1 X_2+\sum_{j=2}^{n-1} X_{j-1} Z_{j} X_{j+1} + X_{n-1} Z_n\Bigr)\ ,
    \nonumber 
\end{align}
of an open chain of $n$ spins,
comprised of single-spin $Z$ terms of magnitude $\lambda_1$, nearest neigbour $XX$ interactions of magnitude $\lambda_2$ and a three-qubit $XZX$ interaction of magnitude $\lambda_3$.
Because of the open boundary conditions, there is a $Z_1X_2$ interaction and an $X_{n-1}Z_n$ interaction of the first and last pair in the chain that resembles the remains of the three-body $XZX$ interactions obtained from truncating such a spin chain.

The properties of the system eigenstates depend on the actual values of the scalar constants $\lambda_1$, $\lambda_2$ and $\lambda_3$ as sketched in Fig.~\ref{fig:labelledtriangle}.
Because an overall prefactor of the Hamiltonian defines only a natural energy scale, but does not change the physics of the model in any other way, the three system parameters can be chosen to adopt the normalization
\be
\sum_j\lambda_j=
\boldsymbol{\lambda}\ ,
\label{eq:scale}
\ee
admitting the two-dimensional representation shown in Fig.~\ref{fig:labelledtriangle}.

\begin{table*}[t]
\centering
\renewcommand{\arraystretch}{1.3}
\begin{tabular} {| c | c | c | c | c |}
\hline
\hspace{.4cm}
elements of Lie algebra
\hspace{.4cm}$\left.\right.$&
\hspace{.3cm}
set
\hspace{.3cm}$\left.\right.$&
\hspace{.3cm}
index range
\hspace{.3cm}$\left.\right.$&
number of elements &
\hspace{.1cm}
row
\hspace{.1cm}$\left.\right.$
\\\hline \hline
$Z_j$ &
$\mathfrak{h}\subset\mathfrak{m}$&
$1\le j\le n$&
$n$&
(i)\\\hline
$X_jZ_{j+1}\hdots Z_{k-1}X_k$
&
$\mathfrak{m}$&
$1\le j<k\le n$&
$n(n-1)/2$&
(ii)\\\hline
$X_jZ_{j+1}\hdots Z_{k-1}Y_k$&
$\mathfrak{k}$&
$1\le j<k\le n$&
$n(n-1)/2$&
(iii)\\\hline
$Y_jZ_{j+1}\hdots Z_{k-1}X_k$&
$\mathfrak{k}$&
$1\le j<k\le n$&
$n(n-1)/2$&
    (iv)\\\hline
$Y_jZ_{j+1}\hdots Z_{k-1}Y_k$&
$\mathfrak{m}$&
$1\le j<k\le n$&
$n(n-1)/2$&
    (v)\\\hline
$Z_1\hdots Z_{j-1}X_j\hspace{1.5cm}$&
$\mathfrak{m}$&
$1\le j\le n$&
$n$&
    (vi)\\\hline
$Z_1\hdots Z_{j-1}Y_j\hspace{1.5cm}$&
$\mathfrak{k}$&
$1\le j\le n$&
$n$&
(vii)\\\hline
$\hspace{1.5cm}X_jZ_{j+1}\hdots Z_{n}$&
$\mathfrak{m}$&
$1\le j\le n$&
$n$&
(viii)\\\hline
$\hspace{1.5cm}Y_jZ_{j+1}\hdots Z_{n}$&
$\mathfrak{k}$&
$1\le j\le n$&
$n$&
(ix)\\\hline
$\mathbf{Z}=Z_1\hdots Z_n$ &  
$\mathfrak{h}\subset\mathfrak{m}$&&
$1$&
(x)\\\hline
\end{tabular}
\caption{
Elements of the Lie algebra generated by the operators in Eq.~\eqref{eq:transLie} (first column) and sets to which the operators belong (second column);
any element of $\mathfrak{h}$ is also element of $\mathfrak{m}$.
The third column depicts the range of the indices of the operators in the first column, and the fourth column depicts the resultant number of operators.\\
With $n$ elements each in rows (i), (vi) and (viii),
$n(n-1)/2$ elements each in rows (ii) and (v)
and one element in row (x), there are $n^2+2n+1$ elements in $\mathfrak{m}$.
With $n(n-1)/2$ elements each in rows (iii) and (iv),
and $n$ elements each in rows (vii) and (ix), there are $n^2+n$ elements in $\mathfrak{m}$.
The full Lie algebra thus has $2n^2+3n+1$ elements.\\
The maximal abelian subset $\mathfrak{h}$ of  $\mathfrak{m}$ (rows (i) and (x)) has $n+1$ elements.
}
\label{tab:Liealgebra}
\end{table*}

For $\lambda_1\gg\lambda_2,\lambda_3$, the single-qubit $Z$ terms dominate and the system is deep in a trivial paramagnetic phase.
For $\lambda_2\gg\lambda_1,\lambda_3$, the two-body $XX$ interactions dominate, and the system is deep in a symmetry-broken Ising-ordered phase. And finally, for $\lambda_3\gg\lambda_1,\lambda_2$, the three-body $XZX$ interactions (together with the two-body boundary terms)
dominate and the system is deep in a symmetry-protected-topological (SPT) phase~\cite{ding2019phase}.
If two of constants $\lambda_j$ are of comparable magnitude, the system shows critical behaviour associated with a quantum phase transition.
The energy gap between the ground state and first excited state is finite for any finite system size, but the gap closes asymptotically with a power-law in the large-system limit. 
The critical and non-critical regions are depicted in the triangular phase diagram (Fig.~\ref{fig:labelledtriangle}), where each corner indicates a dominant phase. Because the locations of quantum phase transitions are not sharply defined, we shade a broad region to indicate their approximate positions. The additional features of Fig.~\ref{fig:labelledtriangle} are discussed in Sec.~\ref{sec:results}.

\subsubsection{System Hamiltonian}

While the single qubit terms $Z_j$ and the interactions $X_jX_{j+1}$ can be realized directly in many platforms of well-controllable qubit registers, the three-body interactions $X_{j-1}Z_{j}X_{j+1}$  generically needs to be realized via effective processes, and thus it cannot be assumed intrinsic to realistic hardware and should not be included in the system Hamiltonian.

In the following, the system Hamiltonian is thus considered to be of the form
\be
H(t)=g\sum _{j=1}^{n-1}X_jX_{j+1}+\sum_{j=1}^{n}h_j(t)Z_j +\sum_{j\in\{1,n\}}h^{(x)}_{j}(t)X_j\ ,
\label{eq:explicitcontrolham}
\ee
with a nearest neighbour $XX$ interaction of constant magnitude $g$, tuneable time-dependent single-spin energies $Z_j$ and additional tuneable $X$ driving of the end spins in the chain.
The interaction constant $g$ defines a natural timescale of any system dynamics beyond single-spin rotations; it also determines a minimal time required to achieve the thermal state preparation, known as the quantum speed limit \cite{deffner2013quantum, deffner2017quantum}.
Modulating the single qubit energies $Z_j$ enables effectively weakening of the $XX$ interactions, similar to the process of dynamical decoupling.
The three-body $XZX$ interactions in Eq.~\eqref{eq:targetinvariant} can be obtained with this system Hamiltonian as effective third-order processes owing to the commutation relation
\be
[X_{j-1}X_j,[X_jX_{j+1},Z_j]]=4X_{j-1}Z_jX_{j+1}\ .
\label{eq:nestedcommutations}
\ee

The system Hamiltonian $H(t)$ (Eq.~\eqref{eq:explicitcontrolham}) thus seems sufficiently elementary to be implemented with suitable time-dependence in practice, but also sufficiently general to induce the dynamics required to emulate the cluster Ising model of Eq.~\eqref{eq:targetinvariant}.

\subsection{Lie algebra and Cartan decomposition}
\label{sec:algebra}

The Lie algebra generated by the operators 
\begin{subequations}
\begin{flalign}
X_1&\ ,\\
X_n&\ , \\
Z_j&\ ,\\
X_{j}&X_{j+1}\ , 
\end{flalign}
\label{eq:transLie}
\end{subequations}
contained in the system Hamiltonian $H(t)$ (Eq.~\eqref{eq:explicitcontrolham})
is a maximal set of linearly independent operators that includes all terms in Eq.~\eqref{eq:transLie}, commutators and nested commutators of these terms.
For example, the operator $Y_1X_2$ is an element of the Lie algebra, since it is obtained via the commutator of the single qubit term $Z_1$ and the two-qubit interaction $X_1X_2$.
The interaction term $Z_1Z_2$, on the other hand is not part of the Lie algebra, since it can not be obtained in terms of nested commutators of the terms in Eq.~\eqref{eq:transLie}.

A Cartan decomposition can be defined with the help of an involution $\Lambda$, {\it i.e.}, a linear map such that $\Lambda^2$ is the identity map and such that $\Lambda([A,B])=[\Lambda(A),\Lambda(B)]$ for any pair of operators $A$ and $B$.
The commutation relations Eq.~\eqref{eq:cartan} are necessarily satisfied, if the elements $k\in\mathfrak{k}$ satisfy the eigenvalue relation $\Lambda(k)=k$ and the elements $m\in\mathfrak{m}$ satisfy the eigenvalue relation $\Lambda(m)=-m$.

For the specific Lie algebra generated by Eq.~\eqref{eq:transLie}, each element is a string of Pauli operators, and a Cartan decomposition can be defined in terms of the transposition $\Lambda(A)=-A^T$.
Since the operator $Y$ satisfies the relation $Y=-Y^T$, while $X$ and $Z$ satisfy the relations $X^T=X$ and $Z^T=Z$,
the operators $k$ have an odd number of factors $Y$, whereas the operators $m$ have an even number of factors $Y$.

Table~\ref{tab:Liealgebra} depicts all the elements of the Lie algebra generated by Eq.~\eqref{eq:transLie} and specifies to which set in the Cartan decomposion each element belongs.
A maximal Abelian subset $\mathfrak{h}\subset\mathfrak{m}$ is given by
\be
\mathfrak{h}=\left\{
Z_1,\hdots,Z_n,{\bf Z}\right\}
\label{eq:seth}
\ee
including all the single-qubit operators $Z_j$ and the $n$-body interaction
\be
{\bf Z}=\prod_{j=1}^nZ_j\ ,
\ee
that will be referred to as {\em parity} in the following.

For the optimization (Eq.~\eqref{eq:optimization}) this means that the initial condition $\Hinit$ is of the form
\be
\Hinit=c_0{\bf Z}+\sum_{j=1}^n c_jZ_j
\label{eq:initialcond}
\ee
with scalars $c_j$ to be optimized over.

The initial state $\varrho^\beta(0)$ for an experimental implementation of the thermal state $\propto\exp(-\beta K_{\bf T})$ following Eq.~\eqref{eq:thermalstatemap} is given by $\varrho^\beta(0)\propto\exp(-\beta \Hinit)$ with the values of the coefficients $c_j$ obtained from the solution of the optimization (Eq.~\eqref{eq:optimization}). 

Moreover, the initial states $\varrho^\beta(0)$ to be prepared are generally diagonal in the computational basis defined by the operators $Z_j$, but the $n$-body interaction term $\mathbf{Z}$ implies that the initial thermal states are generally not product states.
Nevertheless, the states can be prepared rather efficiently as discussed in more detail in Sec.~\eqref{sec:initialstate}.


\subsection{Control results}
\label{sec:results}


\begin{figure}[t]
\centering
\includegraphics[width=\linewidth]{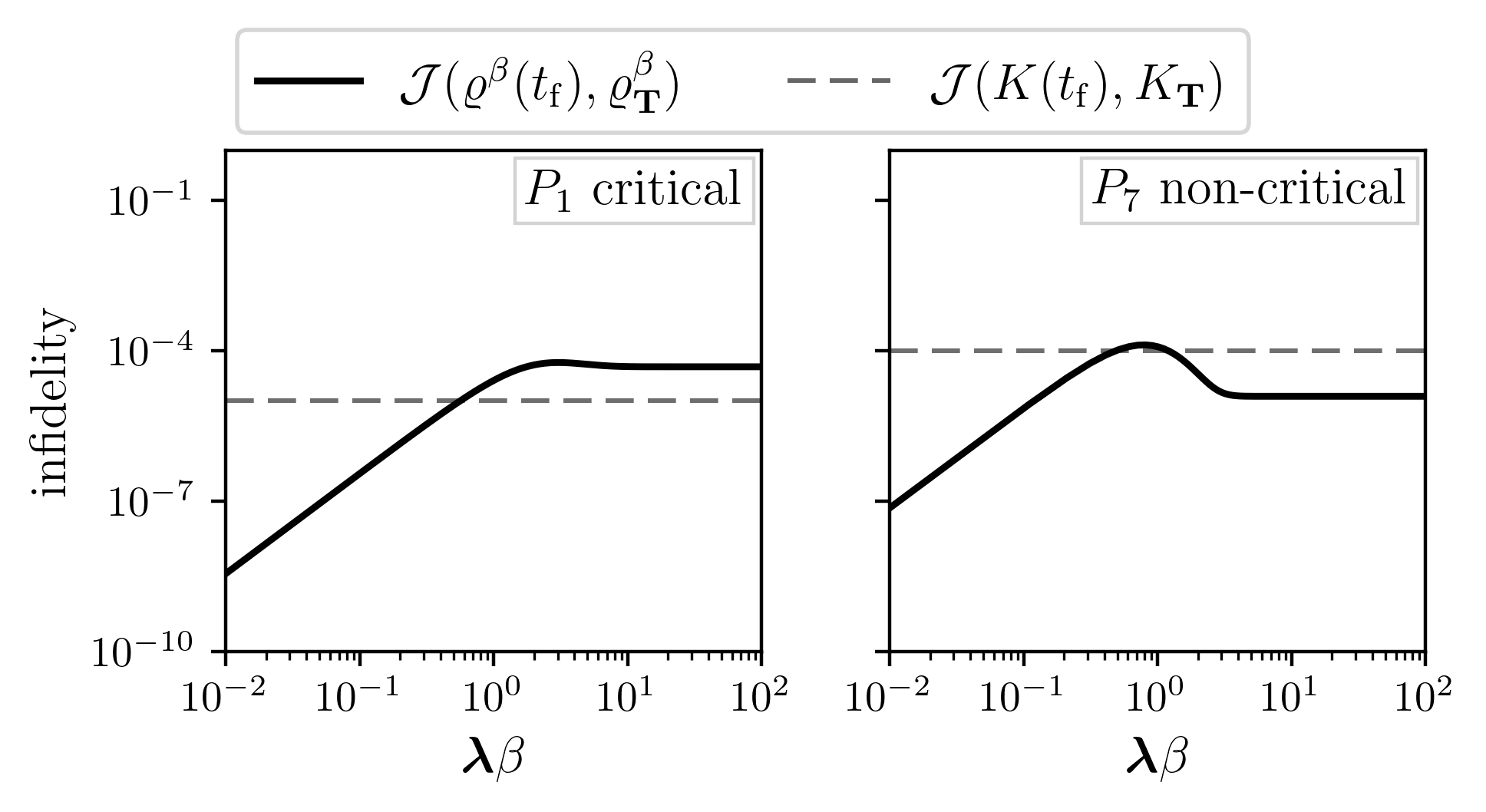}
\caption{
Thermal state infidelity (Eq.~\eqref{eq:thermalinfidelity}) as a function of $\boldsymbol{\lambda}\beta$, with the scale $\boldsymbol{\lambda}$ given in Eq.~\eqref{eq:scale}, for the two exemplary points $P_1$ and $P_7$ depicted in the phase diagram in Fig.~\ref{fig:labelledtriangle} for a system of $n=11$ spins.
The optimal control solution for the thermal state preparation is obtained by optimizing for the operator infidelity, and the final value obtained in the numerical optimization is depicted as a dashed line.\\
The point $P_1$ (left inset) corresponds to critical behaviour with a closing gap; the state fidelity at low temperatures is thus larger than the operator fidelity in contrast to the case of non-critical behaviour depicted in the right inset. In both cases, it is evident that the actual figure of merit (the thermal state infidelity ${\cal J}(\varrho^\beta(t_{\mathrm{f}}),\varrho_\mathbf{T}^\beta)$) remains small even though the proxy (the operator infidelity $\mathcal{J}(K(t_{\mathrm{f}}),\Htarget)$) was optimized.
}
\label{fig:thermalstates}
\end{figure}

\begin{figure}[t]
\centering
\includegraphics[width=\linewidth]{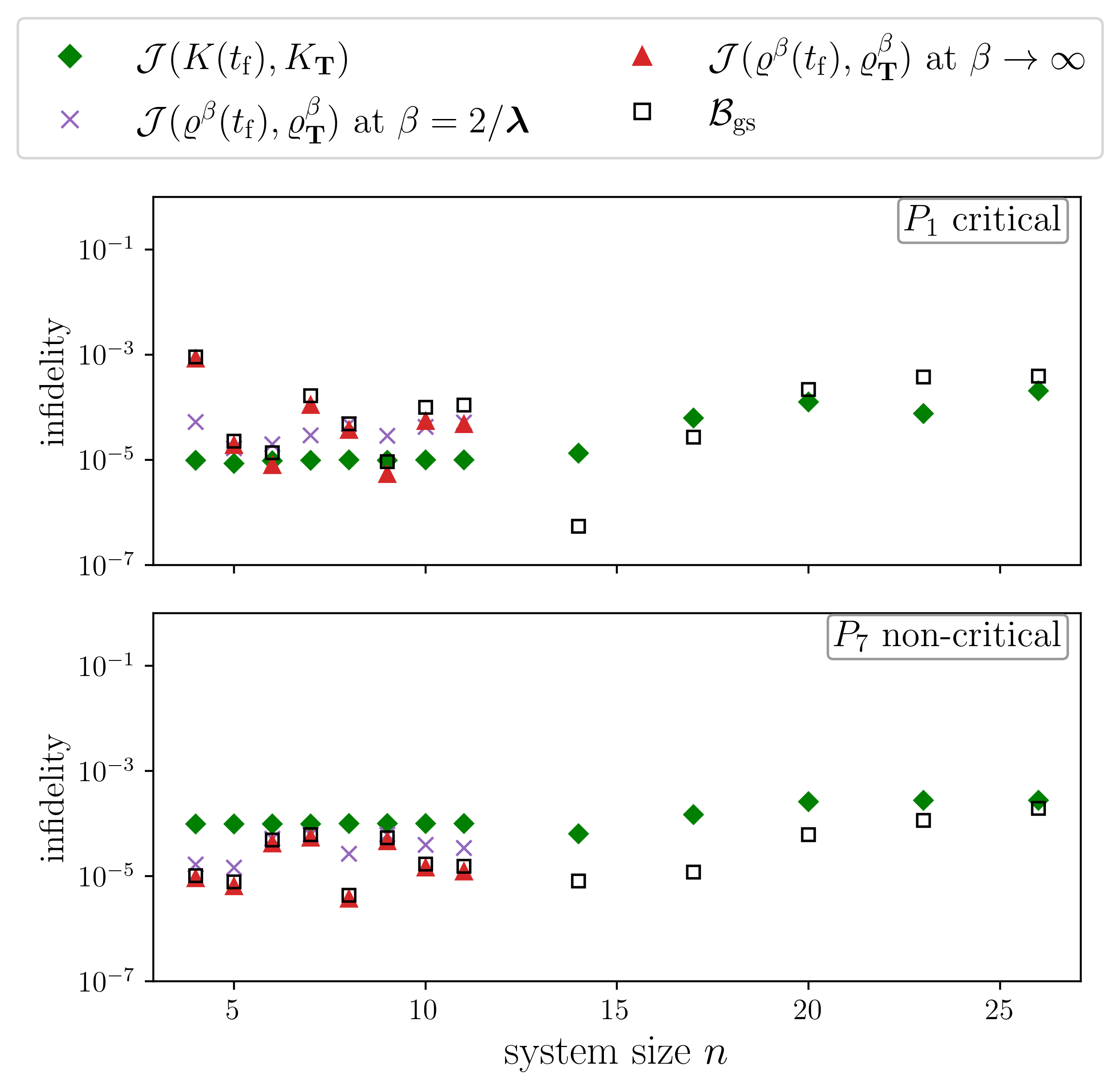}
\caption{
Operator infidelity resultant from the optimization for the two exemplary points $P_1$ and $P_7$ depicted in Fig.~\ref{fig:labelledtriangle} as a function of system size (green diamonds).
State fidelities obtained with the corresponding optimal control solutions are depicted for system sizes up to $n=11$ spins both for the ground state ($\beta\to\infty$, red triangles) and for the finite temperature $\beta=2/\bm{\lambda}$ (purple crosses) with the scale $\bm{\lambda}$ (Eq.~\eqref{eq:scale}) of $K_\mathbf{T}$. The ground state infidelity upper bound $\mathcal{B}_\mathrm{gs}$ (Eq.~\eqref{eq:gsBound}) is depicted by empty square points.
\\
Similar to Fig.~\ref{fig:thermalstates}, state infidelities tend to be larger than operator infidelities in critical regions (top panel) and smaller in non-critical regions (bottom panel).
In both panels, the operator infidelity is a sufficiently good proxy for the state infidelity to be used as control target.
}
\label{fig:opandth}
\end{figure}

In the following, we outline the optimization results depicted by step (i) in Fig.~\ref{fig:scheme}, and for moderate system sizes, we explicitly simulate the thermal state dynamics depicted by step (iii) in Fig.~\ref{fig:scheme}.

With the cluster Ising model defined in Eq.~\eqref{eq:targetinvariant}, the system Hamiltonian defined in Eq.~\eqref{eq:explicitcontrolham}, and the initial condition $K(0)=\Hinit$ restricted to the set $\mathfrak{h}$ in Eq.~\eqref{eq:seth}, the general optimization problem given in Eq.~\eqref{eq:optimization} is completely specified with the definition of the objective function ${\cal J}$ and the parametrization of the time-dependencies of the functions $h_j(t)$ to be optimized.

Because of its practical evaluability, all of the explicit results discussed in the following are based on the infidelity
\begin{equation}
    \mathcal{J}(A,B)=1-\frac{\text{tr}\ AB}{\sqrt{\text{tr}\left(A^2\right)\text{tr}\left(B^2\right)}}
    \label{eq:objfunction}
\end{equation}
as the explicit choice for the objective function $\mathcal{J}(K(\tf),\Htarget)$ in Eq.~\eqref{eq:optimization}.

A brief discussion of subtleties resultant from this choice of objective function, the explicit parametrization of time-dependencies as required for the numerical implementation of the optimization, and the implementation used for this work are provided in App.~\ref{app:suppOpt}.

Explicit control targets are thermal states corresponding to parameters depicted in Fig.~\ref{fig:labelledtriangle} with points $P_5$, $P_6$, and $P_7$ in a regime with one single dominant term in the target Hamiltonian; points $P_2$, $P_3$, and $P_4$ in a regime in which two terms are of comparable magnitude and the third term of much smaller magnitude; and the point $P_1$ in a regime with all the three terms of the Hamiltonian of comparable magnitude.

A vanishing value of the {\em operator infidelity}
\be
\mathcal{J}(K(\tf),\Htarget)
\label{eq:operatorinfidelity}
\ee
verifies that the final operator $K(\tf)$ matches the control target $\Htarget$ exactly. 
Even though an exactly vanishing infidelity indicates that the actual goal -- realization of the desired thermal state -- can be achieved perfectly, minimization of the operator infidelity is not the actual goal of the optimal control problem;
it is rather a substitute problem that is advantageous to implement, but the actual figure of merit of interest is the {\em state infidelity}
\be
{\cal J}(\varrho^\beta(\tf),\varrho_\mathbf{T}^\beta)
\label{eq:thermalinfidelity}
\ee
of the system state $\varrho^\beta(\tf)$ (Eq.~\eqref{eq:thermalstatemap}) at the final time $\tf$ with respect to the target state $\varrho_\mathbf{T}^\beta$ (Eq.~\eqref{eq:targetstate}).
Given the exponential scaling of the Hilbert space, this infidelity can be evaluated in practice only for systems with a moderate number of spins, such as $n=11$, as is the case in Fig.~\ref{fig:thermalstates}. However, owing to the polynomial scaling of the Lie algebra (Tab.~\ref{tab:Liealgebra}), Eq.~\eqref{eq:operatorinfidelity} can be optimized for system sizes beyond what is classically simulable in terms of thermal state dynamics. In such cases quantum hardware is required to explicitly verify the state.

Fig.~\ref{fig:thermalstates} depicts the state fidelity as a function of $\bm{\lambda} \beta$ (with $\bm{\lambda}$ defined in Eq.~\eqref{eq:scale}) for the exemplary cases $P_1$ and $P_7$.
A dashed line depicts the value of the operator infidelity obtained for this optimization.
The exemplary cases $P_1$ and $P_7$ include one critical and one non-critical point in the phase diagram Fig.~\ref{fig:labelledtriangle}.
Corresponding data for $P_2$ to $P_6$ is shown in Fig.~\ref{fig:thermalstatessupp} in App.~\ref{app:suppSol} to provide evidence for qualitatively similar behaviour of all critical points and of all non-critical points.

The ground state infidelity ({\it i.e.} the limit $\beta\to\infty$) is larger than the operator infidelity for $P_1$, but it is smaller than the operator infidelity for $P_7$. 
This can be attributed to the fact that the control target $K_{\bf T}$ has a sizeable gap for $P_7$, whereas the gap for $P_1$ is closing as a power law in the total number of spins.
Since the accuracy of the ground state fidelity is bounded by the operator fidelity with a bound that depends on the gap~\cite{orozco2024quantum} it is expected that control targets with a small gap require lower operator infidelities to reach a given ground state infidelity than targets with a larger gap.

At sufficiently high temperatures, the thermal state infidelities decrease as the thermal state approaches full degeneracy; here, imperfections in the state transfer of individual eigenstates are increasingly averaged out by the nearly uniform populations.
At in-between temperatures, a sizeable number of excited states contributes to the thermal state infidelity, but the thermal states are not close to degeneracies.
The infidelities for such thermal states can exceed ground state infidelity, and this is typically the case for states in non-critical regions in which low ground state infidelities are obtained.

Since the comparison between operator infidelity and thermal state infidelity depends also on the system size, 
Fig.~\ref{fig:opandth} 
depicts infidelities achieved with optimized protocols for the points $P_1$ and $P_7$ as a function of system size $n$
(corresponding data for $P_2$ to $P_6$ is shown in Fig.~\ref{fig:suppopandth} in App.~\ref{app:suppSol}).
Thermal state infidelities with the intermediate temperature $\beta=2/\bm{\lambda}$ are depicted for system sizes up to $n=11$.
As seen in Fig.~\ref{fig:thermalstates}, most prominently for the non-critical point, this temperature typically represents a worst-case regime for the thermal state infidelity.
Even in this regime, the thermal state infidelities are typically within an order of magnitude higher than the corresponding operator infidelity for targets in critical regimes.
For non-critical targets, the thermal state infidelity is generally within an order of magnitude lower than the corresponding operator infidelity. 
For targets in the critical region, the ground state infidelity -- {\it i.e.} the thermal state infidelity in the limit $\beta\to \infty$ -- commonly represents the worst-case thermal state infidelity due to the closing ground state energy gap.
This is particularly evident in Fig.~\ref{fig:suppopandth} in App.~\ref{app:suppSol}.
Like the intermediate-temperature state fidelity, this quantity is infeasible to evaluate for large system sizes ($n\ge 14$). 
Therefore, it is instructive to introduce an upper bound for the ground state infidelity that can be evaluated even when explicit classical simulation is infeasible. 
Proven in \cite{orozco2024quantum}, the upper bound is given by
\begin{equation}
    \mathcal{B}_\mathrm{gs} = \frac{\bra{\Psi(\tf)}\Htarget\ket{\Psi (\tf)}-E_0}{E_1-E_0}\ ,
    \label{eq:gsBound}
\end{equation}
where $E_0$ and $E_1$ are the ground state and first excited state energies of $\Htarget$ respectively, and $\ket{\Psi (\tf)}$ is the ground state of the evolved operator $K(\tf)$. The expectation value $\bra{\Psi(\tf)}\Htarget\ket{\Psi (\tf)}$ can be evaluated implicitly, without explicit construction of $\Htarget$ and the full statevector $\ket{\Psi (\tf)}$~\cite{orozco2024quantum}.

Strikingly, in Fig.~\ref{fig:opandth}, it appears that the bound tightens with respect to the operator infidelity as the system size increases.
For the critical example, $P_1$, it appears that the bound is substantially lower for $n=17$, and especially $n=14$.
However, this indicates the existence of an optimal control solution with particularly good ground state infidelity.

Crucially, Figs.~\ref{fig:thermalstates} and \ref{fig:opandth} demonstrate that thermal states in the critical region can be prepared with high-fidelity using the same experimental resources--like evolution time $\tf$--required for non-critical regions.

All the infidelities shown in Figs.~\ref{fig:thermalstates} and \ref{fig:opandth} are finite, but sufficiently small to realize thermal state preparation with accuracies compatible with existing or near-term hardware.
Fundamentally, the achievable value of infidelity is limited by the duration $\tf$ of the controlled dynamics, because the interaction constant $g$ in the system Hamiltonian $H(t)$ (Eq.~\ref{eq:explicitcontrolham}) defines a fundamental minimal time scale (quantum speed limit) for any entangling dynamics. 

While optimal control solutions can overcome restrictions of other existing strategies, such as adiabaticity, the restriction imposed by the quantum speed limit is an intrinsic system property that can not be overcome simply in terms of sufficiently strong or sophisticated driving of single-qubit dynamics.

As such, one would expect that there is a minimal duration $t_\text{min}$ for the evolution time $\tf$ that is required for preparation protocols with vanishing infidelities or, in practice, values of infidelity that are limited only by the accuracy of the numerical opimization.

\begin{figure}[t]
\centering
\includegraphics[width=\linewidth]{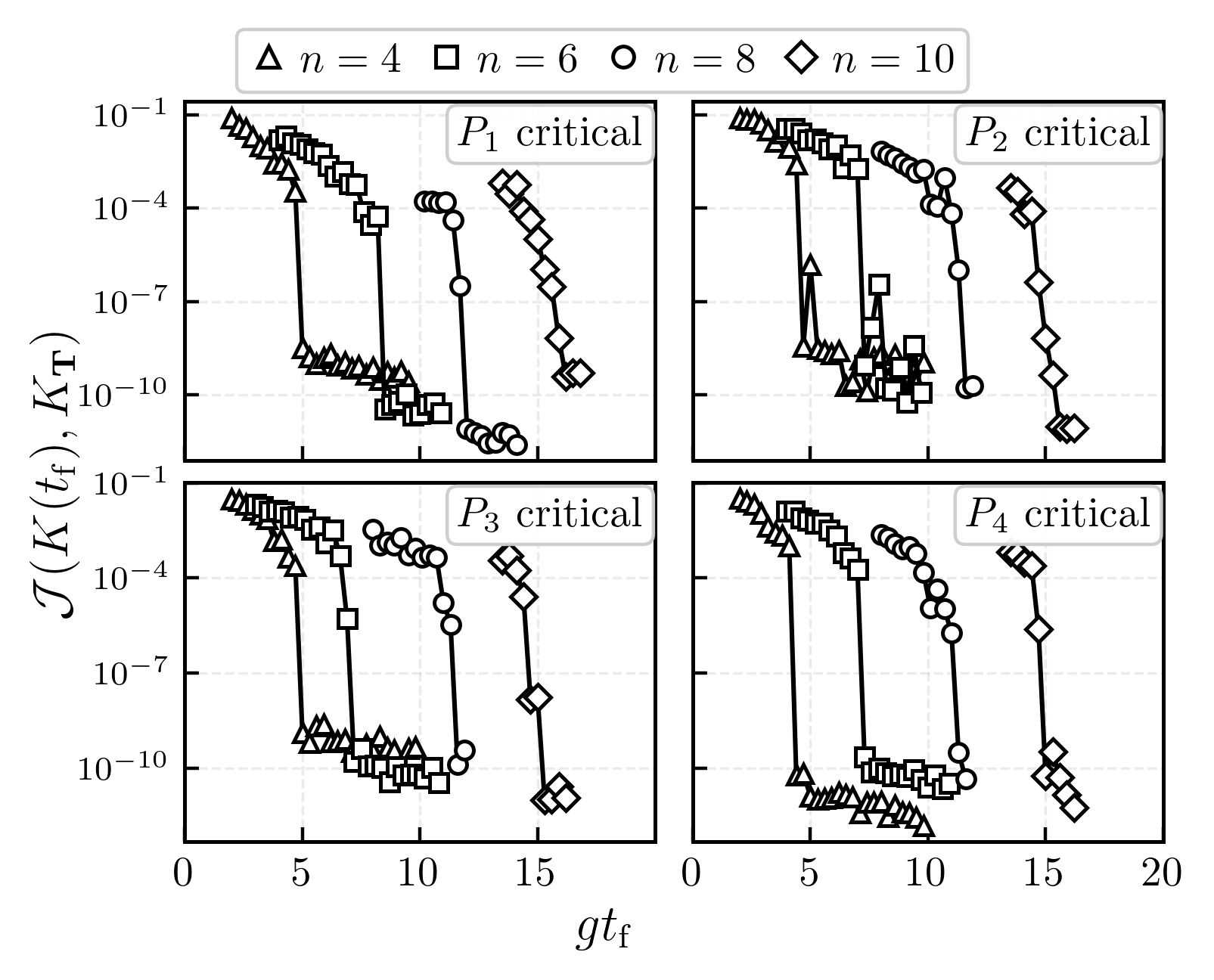}
\caption{Best operator infidelity obtained as result of the optimization as a function of the evolution time $g \tf$, where $g$ is the interaction constant in the system Hamiltonian $H(t)$ (Eq.~\eqref{eq:explicitcontrolham}).
There is a well-defined drop in the infidelity, suggesting that there is a minimal duration of system dynamics required to realize the thermal state preparation.
This minimal duration appears independent of the specific target within the cluster Ising model, and it seems to be growing only moderately with the system size $n$.}
\label{fig:speedlimits}
\end{figure}

Fig.~\ref{fig:speedlimits} depicts values of the operator infidelity as a function of the evolution time for various system sizes for the critical points $P_1$ to $P_4$ (Fig.~\ref{fig:labelledtriangle}).
(Data for the noncritical points $P_5$ to $P_7$ is shown in Fig.~\ref{fig:speedlimitssupp} in App.~\ref{app:suppSol}). 
Apart from numerical noise caused by the common limited reliability of numerical optimizations, the operator infidelity decays monotonically with increasing evolution time $\tf$.
This decay is rather weak for sufficiently short times, but there is a narrow time-window, in which we observe a very pronounced drop of infidelity, down to a value that is limited by the numerical accuracy of the optimization.

The clearly pronounced sudden drop in operator infidelity suggests that the optimization is indeed able to find close-to-perfect control solutions, as soon as the evolution time $\tf$ is sufficiently long to admit such solutions.

While the present approach allows us to overcome the exponential growth of the numerical effort to simulate the system dynamics, the effort to optimize for the time-dependent system Hamiltonian $H(t)$ grows with the system size $n$ ({\it i.e.,} Eq.~\eqref{eq:optimization} requires the optimization of $\mathcal{O}(n^2)$ parameters).
In contrast to the simulation of the system dynamics that mostly requires computational memory (spatial complexity), the optimization requires the ability to try many different time-dependencies; since this can be done sequentially, this implies time-complexity, {\it i.e.,} a cost dominated by the total runtime needed to explore many candidate time-dependencies rather than by memory usage.
As such, one can always attempt to find a better time-dependence than the best found so far, but in practice one will need to accept a solution with some finite accuracy.
As seen in Figs.~\ref{fig:thermalstates} and \ref{fig:opandth}, operator infidelities of $\mathcal{O}(10^{-4})$ can be sufficient to design practical protocols for thermal state preparation, but an unambiguous identification of the quantum speed limit requires substantially lower infidelities.
The numerical assessment of this time-scale is therefore limited to at most $n=10$ spins.

\begin{figure}[t]
\centering
\includegraphics[width=\linewidth]{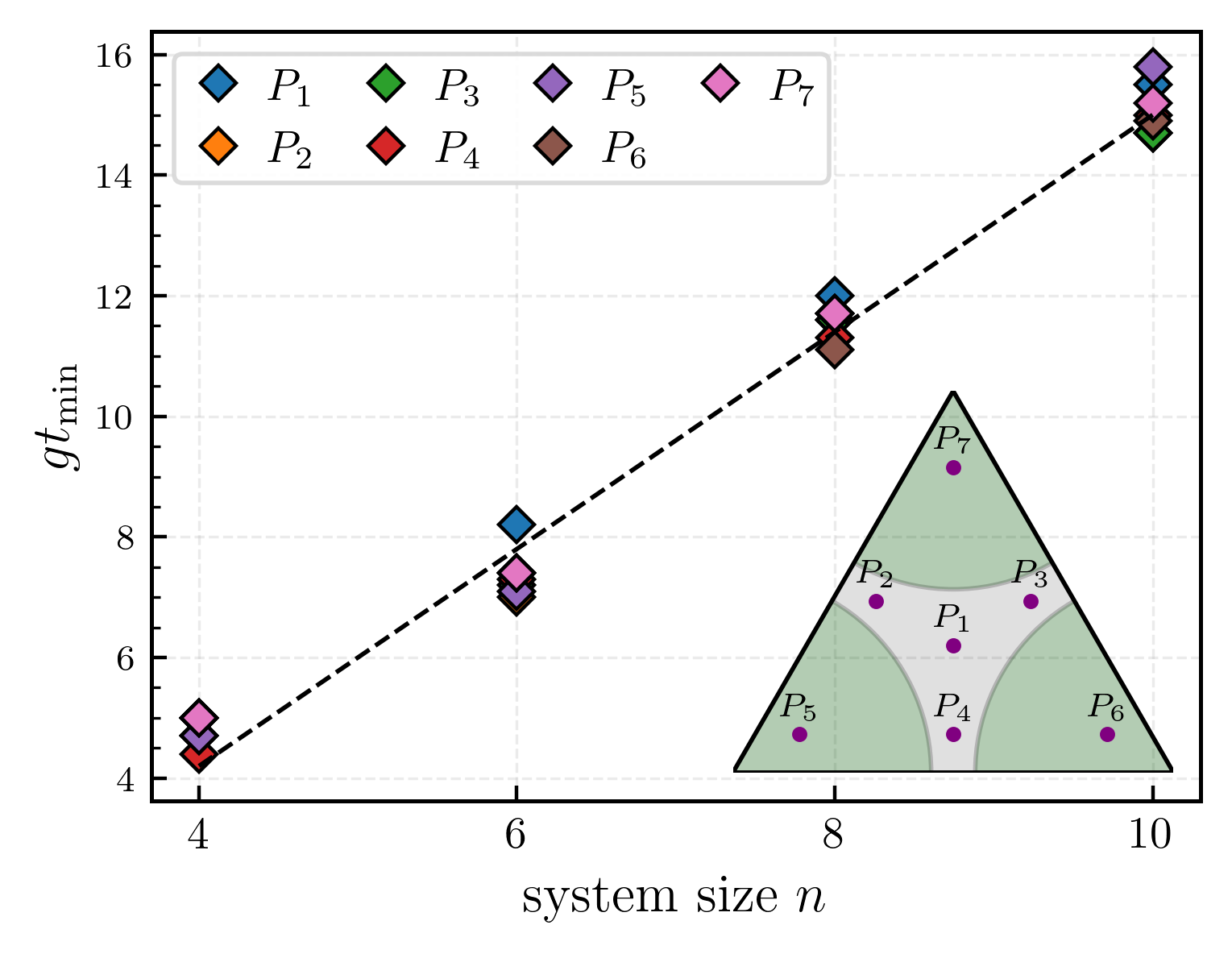}
\caption{The quantum speed limit duration, $gt_\text{min}$, where $g$ is the interaction constant in the system Hamiltonian $H(t)$ (Eq.~\eqref{eq:explicitcontrolham}), as a function of system size $n$. $t_\text{min}$ is identified as the clear drop in operator infidelity shown in Fig.~\ref{fig:speedlimits} and Fig.~\ref{fig:speedlimitssupp} in the supplementary solutions. The data suggests a linear relationship between $t_\text{min}$ and system size, regardless of phase diagram location (Fig.~\ref{fig:labelledtriangle}), which is a clear indication of non-adiabaticity. The phase diagram (Fig.~\ref{fig:labelledtriangle}) is included as a visual aid, along with a dashed linear reference line.
}
\label{fig:tqslvsN}
\end{figure}

For the practically accessible system sizes, however, Fig.~\ref{fig:speedlimits} (and Fig.~\ref{fig:speedlimitssupp} in App.~\ref{app:suppSol}) allows us to read off the system-size dependent quantum speed limit rather accurately.
This duration is depicted for all the points $P_1$ to $P_7$ in Fig.~\ref{fig:tqslvsN} as a function of the system size $n$.
Strikingly, the differences of the minimal duration obtained for different points $P_j$ at a given system size are within the accuracy to which this time can be read off of Figs.~\ref{fig:speedlimits} and \ref{fig:speedlimitssupp}.
There is thus no significantly longer duration required for the preparation of states in critical regimes than for non-critical ones.
This contradicts the common perception that the preparation of critical states is limited by the time-scale imposed by the closing gap.
This perception, however, is based on adiabatic dynamics.
The results in Fig.~\ref{fig:tqslvsN} are thus not in contradiction with the common perception, but they are an indication of the non-adiabatic character of the control solutions that can be found with the present approach.

To the extent that Fig.~\ref{fig:tqslvsN} allows for an identification of a scaling relation, the minimal time required for the state preparation appears to grow linearly with the system size.
This is consistent with the scaling obtained for the preparation of the ground state of the cluster Ising Hamiltonian with $\lambda_1=\lambda_2=0$ in Eq.~\eqref{eq:targetinvariant}~\cite{orozco2024quantum}.
This scaling also highlights that the state preparation is not relying on an adiabatic approximation, that would yield a scaling with the closing gap.

\subsection{Explicit initial thermal state construction}
\label{sec:initialstate}

In addition to the explicit control solutions constructed to assess the infidelities discussed in Sec.~\ref{sec:results} above, an implementation on a quantum device also needs a procedure to prepare the initial thermal state $\varrho^\beta(0)$, as seen in Eq.~\eqref{eq:thermalstatemap}. This part of the broader method is depicted as step (ii) in Fig.~\ref{fig:scheme}.

The construction of optimal control solutions in Sec.~\ref{sec:results} is in the spirit of an analog quantum simulation with a Hamiltonian (Eq.~\eqref{eq:explicitcontrolham}) that has a constant interaction and additional single-spin driving that is being applied while the spins are interacting.
The framework of optimization is, however, also directly applicable to a digital quantum simulation with two-qubit $XX$-gates and single-qubit gates that are being applied sequentially.
While the present framework to optimize the thermal state preparation is applicable to both analog and digital quantum simulations, the initial state preparation is likely to be different in the two approaches.
The following discussion of initial thermal state preparation thus includes two different approaches.

The first approach realizes the desired initial thermal states in terms of an average over randomly selected pure states.
This approach does not require the application of any entangling gates, and is thus suitable for analog quantum simulations.
The second approach is defined in terms of a gate sequence that realizes a pure state with the occupations that are consistent with the desired mixed intial state, and that turns this coherent superposition into an incoherent mixture with additional entangling gates that involve additional qubits that play the role of an environment.
This approach does require the capabilities of a digital quantum simulator.

\subsubsection{The initial state}

As discussed in Sec.~\ref{sec:algebra}, the initial condition $\Hinit$
is of the form
\be
\Hinit=c_0{\bf Z}+\sum_{j=1}^nc_jZ_j
\label{eq:Zinit}
\ee
with the weights $c_i$ of the individual terms determined as solutions of the optimal control problem.
The initial condition for an actual implementation is then a thermal state
\be
\varrho^\beta(0)=\frac{\exp(-\beta \Hinit)}{\tr\exp(-\beta \Hinit)} .
\label{eq:initthermal}
\ee
with the initial condition $\Hinit$ (Eq.~\eqref{eq:Zinit}) as the parent Hamiltonian.

Without the parity ${\bf Z}$, {\it i.e.} the $n$-body interaction, the initial condition $\varrho^\beta(0)$ would be a product of single-qubit thermal states that is rather straightforward to prepare.
The following discussion will thus mostly focus on the aspects required to prepare a state that is not a product state.

\subsubsection{Random sampling}

Any mixed state can be realized in terms of an ensemble of pure states with associated probabilities.
Such probabilities can easily be constructed for the initial $\varrho^\beta(0)$ (Eq.~\eqref{eq:initthermal}),
but in order to realize the state preparation in an experiment, it is important that one can sample efficiently from the exponentially large ensemble.


A precondition for finding such a sampling procedure is the construction of the probabilities associated with each pure state in the ensemble.
It is natural to take these pure states to be the eigenstates of the initial operator $\Hinit$ (Eq.~\eqref{eq:Zinit}), {\it i.e.} the product states
\be
\ket{\vec z_n}=\ket{z_1}\otimes\hdots\otimes\ket{z_n}
\ee
of single-spin $Z_j$ eigenstates.

For any traceless Hamiltonian with eigenvalues $\pm c$, the probability $p(z)$ for a given pure state with eigenvalue $zc$ and $z\in\{-1,1\}$ in an ensemble representing a thermal state is given by
\be
p(z)=\frac{1}{n_d}\frac{\exp(-\beta cz)}{\exp(-\beta c)+\exp(\beta c)}\ ,
\ee
where $n_d$ is the degree of degeneracy of the eigenvalues.
Given that the quantum number $z$ adopts only the values $\pm 1$, this can be written equivalently as
\be
p(z)=\frac{1}{n_d}Q(m,z)\ ,\ \mbox{with }\
Q(m,z):=\frac{1-mz}{2}\ ,
\ee
where $m=\tanh(\beta c)$.

Since the operator $\Hinit$ (Eq.~\eqref{eq:Zinit}) is a sum of mutually commuting terms, the probability for a given state $\ket{\vec z_n}$ in the ensemble representing the thermal state $\varrho^\beta(0)$ can be written as
\be
p(\vec z_n)=
\frac{1}{N_\beta}
\prod_{j=0}^n
Q(m_j,z_j)\ ,
\label{eq:probability}
\ee
with $m_i=\tanh(\beta c_i)$ and $z_0=\prod_{j=1}^nz_j$;
this is essentially the product of the probabilities resultant from each term in the Hamiltonian,
but since the quantum number $z_0$ associated with the operator ${\bf Z}$ is not an independent variable, the normalization $N_\beta$ is not just the product of the normalizations of the individual probabilities.

Leaving aside that the normalization factor $N_\beta$ is still undetermined,
one can construct the marginal probabilities $p(\vec z_{j-1})=\sum_{z_{j}=\pm 1}p(\vec z_j)$, and one obtains
\begin{equation}
p(\vec z_j)=
\frac{1}{N_\beta}
Q\left(m_0^{(j)},\prod_{i=1}^jz_i\right)
\prod_{i=1}^{j}Q(m_i,z_i)
\label{eq:marginalprob}
\end{equation}
with
\be
m_0^{(j)}=
\left(-1\right)^{n-j}m_0
\prod _{i=j+1}^nm_i\ .
\ee
The condition $\sum_{z_1=\pm 1}p(\vec z_1)=1$ finally determines the value
\begin{equation}
N_\beta=\frac{1}{2}
\left(1-\left(-1\right)^n \prod_{i=0}^nm_i\right)
\end{equation}
of the normalization constant $N_\beta$.

With the marginal probabilities $p_j(\vec z_j)$ (Eq.~\eqref{eq:marginalprob}) for the states of the first $j$ spins,
one also obtains the conditional probabilities
\be
p(z_{j+1}|\vec z_j)=\frac{p(\vec z_{j+1})}{p(\vec z_{j})}
\label{eq:condprob}
\ee
for the spin state $\ket{z_{j+1}}$ given a state $\ket{\vec z_j}$ of the first $j$ spins,
and these conditional probabilities give access to an efficient sampling procedure.

In order to pick a given state $\ket{\vec z_n}$ with the appropriate probability, one can pick a state $\ket{z_1}$
of the first spin with the probability $p(\vec z_1)$ given in Eq.~\eqref{eq:marginalprob}. 
The probability for a state $\ket{z_2}$ of the second spin is given by the conditional probability $p(z_2|\vec z_1)$ (Eq.~\eqref{eq:condprob}), which can be explicitly evaluated because a state $\ket{z_1}$ for the first spin is selected.
This autoregressive sampling follows through to the last spin.

This procedure allows one to sample pure states $\ket{\vec z_n}$ from the probabilities in Eq.~\eqref{eq:probability}. An average over samples created in this way realizes the desired thermal state $\varrho^\beta(0)$.

\subsubsection{Gate circuit}

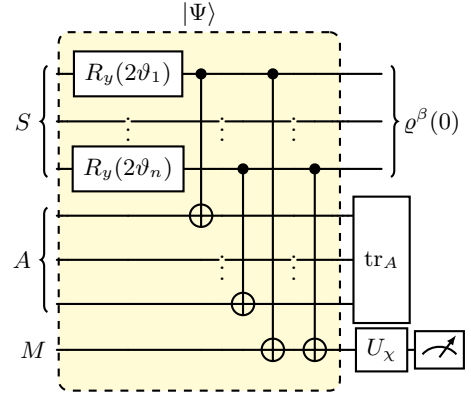
\begin{figure}[t!]
\centering
\begin{quantikz}[column sep=2pt, row sep=2pt]
\lstick[3]{$S$} & & & \gate{R_y(2\vartheta_1)} \gategroup[7,steps=7,style={dashed,rounded
corners, fill=yellow!20, inner
xsep=2pt},background,label style={label
position=above,anchor=north,yshift=+0.3cm}]{$\ket{\Psi}$}  &  \ctrl{3} &  &  & \ctrl{6}  &  & & & &  & &  \rstick[3]{$\varrho^\beta(0)$} \\
 & & & \push{\vdots} &  &   \push{\vdots} &  & & \push{\vdots} &  & & & & &   \\
 & & & \gate{R_y(2\vartheta_n)} &  &  & \ctrl{3} & & & \ctrl{4} &  &  & & &  \\
\lstick[3]{$A$} & & &  & \targ{} &  & & & &  & &  & & & \gate[3]{\text{tr}_A} \\
 & & &  &  & \push{\vdots} &  & & \push{\vdots} & & & & & & \\
 & & &  &  &  & \targ{} &  &  & & & &  & & \\
\lstick[1]{$M$} & &  &  &  &  & & \targ{} &  & \targ{} & &  & & & \gate{U_\chi} & \meter{}
\end{quantikz}
\caption{Quantum circuit diagram to realize the state in Eq.~\eqref{eq:thermalstate}.
$S$ and $A$ are $n$ qubit registers, representing system and auxiliary Hilbert spaces, and $M$ is a single qubit to encode parity.
All registers are initialised in $\ket{0}$.
The gate sequence in the yellow box generates the pure state $\ket{\Psi}$, after which discarding the auxiliary system yields, $\text{tr}_A\ket{\Psi}\bra{\Psi}=\varrho_F^\beta$ (Eq.~\eqref{eq:thermalstateparity}).
A unitary $U_\chi=\ket{0}\bra{\chi}+\ket{1}\bra{\chi_\perp}$ applied on qubit $M$, followed by a $Z$-measurement, yields the desired thermal state (Eq.~\eqref{eq:thermalstate}) in register $S$ , where $\left\{\ket{\chi},\ket{\chi_\perp}\right\}$ is an orthonormal basis.
The quantum circuit requires $n$ single qubit $Y$ rotations (with specific angles given in Eq.~\eqref{eq:RyAngles}), and $2n$ CNOTs, where the first $n$ CNOTs can be applied in a single unit of depth.
} 
\label{fig:purificationcirc}
\end{figure}

Alternatively, instead of the classical sampling discussed above, the initial thermal state $\varrho^\beta(0)$ can be prepared by realizing system-environment entanglement with a sequence of gates, akin to the preparation of thermofield double states \cite{cottrell2019build, wu2019variational}. Although the initial condition $\Hinit$ (Eq.~\eqref{eq:Zinit}) includes the $Z$-string $\mathbf{Z}$, so $\varrho^\beta(0)$ cannot be prepared as the product of single-spin mixed states, it is nevertheless instructive to first follow the preparation of single-spin mixed states.

This can be done with the system-spin prepared in the state $\ket{0}$, a $Y$-rotation by an angle determined by the temperature, and a CNOT gate with the system spin as control qubit and an auxilary spin as target qubit to realize the system-environment entanglement.
With $n$ system spins and $n$ corresponding auxilary spins, this allows one to realize any thermal state $\varrho_{\text{NI}}^\beta$ with a non-interacting initial condition $\Hinit=\sum_{j=1}^nc_jZ_j$, {\it i.e.} $c_0=0$ in Eq.~\eqref{eq:Zinit}.

In the general case of a finite parity component in the initial condition ({\it i.e. $c_0\neq 0$}), the initial thermal state to be prepared is of the form
\be
\varrho^\beta(0)\propto\varrho_{\text{NI}}^\beta e^{-\beta c_0{\bf Z}}\ .
\ee
With the identity
\be
e^{-\beta c_0{\bf Z}}=
e^{-\beta c_0}P_++e^{\beta c_0}P_-\ ,
\ee
in term of the projectors $P_{\pm}=\frac{1}{2}(\um\pm{\bf Z})$ onto positive and negative parity,
and commutativity $[{\bf Z},\varrho_{\text{NI}}^\beta]=0$, this reduces to
\be
\varrho^\beta(0)\propto
e^{-\beta c_0}P_+\varrho_{\text{NI}}^\beta P_++e^{\beta c_0}P_-\varrho_{\text{NI}}^\beta P_-\ .
\label{eq:thermalstate}
\ee
The desired state is thus a weighted, incoherent sum of the two parity components $P_+\varrho_{\text{NI}}^\beta P_+$ and $P_-\varrho_{\text{NI}}^\beta P_-$ of the non-interacting state $\varrho_{\text{NI}}^\beta$.

In order to realise such a mixture in a gate circuit, it is helpful to introduce an additional spin $M$,
and to apply $n$ CNOT gates with the individual system spins as control qubits and the spin $M$ as target.

Starting with $\varrho_{\text{NI}}^\beta\otimes\ket{0}\bra{0}_M$, this yields the state
\bqa
\varrho_F^\beta&=&\varrho_{\text{NI}}^\beta P_+\otimes\ket{0}\bra{0}_M+\varrho_{\text{NI}}^\beta P_-\otimes\ket{1}\bra{1}_M\\
&=&\frac{1}{2}\left(\varrho_{\text{NI}}^\beta \otimes\um_M+\varrho_{\text{NI}}^\beta {\bf Z}\otimes Z_M\right)\ . \label{eq:thermalstateparity}
\eqa
A subsequent measurement on spin $M$ in a basis including the state
\be
\ket{\chi}=\frac{
e^{-\frac{\beta c_0}{2}}
\ket{0}+
e^{\frac{\beta c_0}{2}}\ket{1}}{\sqrt{2\cosh(\beta c_0)}}
\ee
yields the desired thermal state (Eq.~\eqref{eq:thermalstate}) conditioned on the projection of the spin $M$ onto the state $\ket{\chi}$.

The success probability $P_s$ of this protocol is given by the norm of the projected state, {\it i.e.}
\be
P_s=\tr\bra{\chi}\varrho_F\ket{\chi}=\frac{1}{2}\left(1+(\tr\varrho_{\text{NI}}^\beta{\bf Z})\bra{\chi}Z\ket{\chi}\right)\ .
\ee
With $\tr\varrho_{\text{NI}}^\beta{\bf Z}=(-1)^n\prod_{i=1}^n\tanh(\beta c_i)$ and $\bra{\chi}Z\ket{\chi}=-\tanh(\beta c_0)$, this reduces to
\be
P_s=\frac{1}{2}\left(
1-(-1)^n\prod_{i=0}^n\tanh(\beta c_i)\right)\ .
\ee


In the high temperature limit, the success probability approaches the value of $1/2$.
In the low temperature limit,
the success probability is close to unity if the non-interacting part $\sum_{j=1}^nc_jZ_j$ has the correct parity to also be a ground state of $c_0{\bf Z}$.
In the opposite case the success probability can also be too low to be practical.
In this, one can use the deterministic sampling methods instead.



Fig.~\ref{fig:purificationcirc} depicts a circuit diagram to realize this state preparation.
It consists of $n$ qubits labelled $S$ to realize the system spins, $n$ qubits labelled $A$ to serve as auxiliary spins and one additional qubit labelled $M$ required to realize the Boltzmann factor with the total parity ${\bf Z}$.

All system qubits are initialized in their state $\ket{0}$, and single qubit rotations $\exp(- i\vartheta_jY_j)$ with the angles $\vartheta_j$, such that
\begin{equation}
    \cos \vartheta _j=\sqrt{\frac{1-m_j}{2}}\quad \sin \vartheta _j=\sqrt{\frac{1+m_j}{2}},
    \label{eq:RyAngles}
\end{equation}
with $m_i=\tanh(\beta c_i)$ on each of the qubits $S$ ensure that these qubits have the occupations that are consistent with the single-spin thermal state $\propto\exp(-\beta c_jZ_j)$.

A subsequent series of CNOT gates with qubits $S$ as control qubits and qubits $A$ as target qubits ensures that the reduced density matrix of the system of qubits $S$ is given by an incoherent mixture and not by a coherent superposition.

A further series of CNOT gates with the qubits $S$ as control qubits and the qubit $M$ as target qubit is a first step in order to adjust the occupations of the individual components in this mixture to take into account the parity term $\propto\exp(-\beta c_0{\bf Z})$.
After a subsequent $Y$ rotation $\exp (i\phi Y)$ with $\tan(\phi)=e^{\beta c_0}$ on qubit $M$ (equivalent to the application of the unitary $U_\chi$ depicted in Fig.~\ref{fig:purificationcirc}),
projection onto the state $\ket{0}$ upon a readout of qubit $M$, leaves the reduced system of qubits $S$ in the desired thermal state including the parity contribution.

\section{Discussion and outlook}

The present approach enables quantum simulations that require both the realization of effective processes beyond native interactions and state preparation beyond the zero-temperature limit.
As we demonstrate using the cluster Ising model thermal state as the control target, our method works reliably both throughout the entire phase diagram and across all temperature regimes. Unlike previous approaches, our results show that the thermal state evolution duration $\tf$ is strikingly unaffected by aspects like criticality or temperature.
While the cluster Ising model is an example to illustrate the workings of the present framework, the methodology is directly applicable to any thermal state of a Hamiltonian within the Lie algebra discussed in this paper or any other Lie algebra with suitable scaling -- some examples can be found in \cite{wiersema2024classification}.

Aspects like disorder in the system Hamiltonian $H(t)$ and in the target Hamiltonian $K_{\bf T}$ are within the scope of a given Lie algebra~\cite{stefanescu2024robust} and can thus be easily taken into account in the construction of explicit solutions.
Also, the restriction to thermal states in this work is due to their practical importance, but the framework is directly applicable also to mixed states that are not of Gibbs-form.

Any problem that does not fit into the framework of a small Lie algebra does require numerical approximations in order to benefit from the methodology laid out here.
Additional terms that exceed the utilized Lie algebra can be taken into account perturbatively~\cite{stefanescu2024robust} without jeopardizing the numerical efficiency that lies at the core of the present framework.

While the explicit optimizations presented here are in the spirit of an analogue quantum simulation, the formalism can equally well be applied to the optimization of gate parameters for a digital quantum simulation.
In such a case, there is no intrinsic interaction constant to define a fundamental time scale, and the figure of merit replacing the overall system evolution time could be the number of entangling gates that need to be applied to a pair of qubits before infidelities on the level of numerical accuracy can be reached.

Practical considerations such as addressability of individual qubits or spectral properties of the driving functions can be taken into account in the optimization.
The explicit control solutions presented here, assume that each of the qubits can be driven individually, but if an actual device cannot achieve perfect addressing, or even only global control, then the framework can be applied to a system Hamiltonian that respects such a restriction.

With the flexibility to accommodate practical requirements and to realize challenging quantum simulations in a fashion that requires minimal resources from a quantum device, the present approach allows us to push the envelope of quantum simulation further towards practical use-cases.

\section*{Acknowledgements}

This work was supported by the U.K. Engineering and Physical Sciences Research Council (EPSRC DTP - EP/W524323/1). 
MB and PMS were funded by the European Union (ERC, QuSimCtrl, 101113633). Views and opinions expressed are however those of the authors only and do not necessarily reflect those of the European Union or the European Research Council Executive Agency. Neither the European Union nor the granting authority can be held responsible for them.
Numerical simulations and optimization routines were performed on the Imperial HPC cluster.

\section*{Data availability}
The optimal control solutions used in this paper are available without restriction \cite{van_lomwel_data}.

\appendix
\section{Supplementary control solutions} \label{app:suppSol}

\begin{figure}[t!]
\centering
\includegraphics[width=\linewidth]{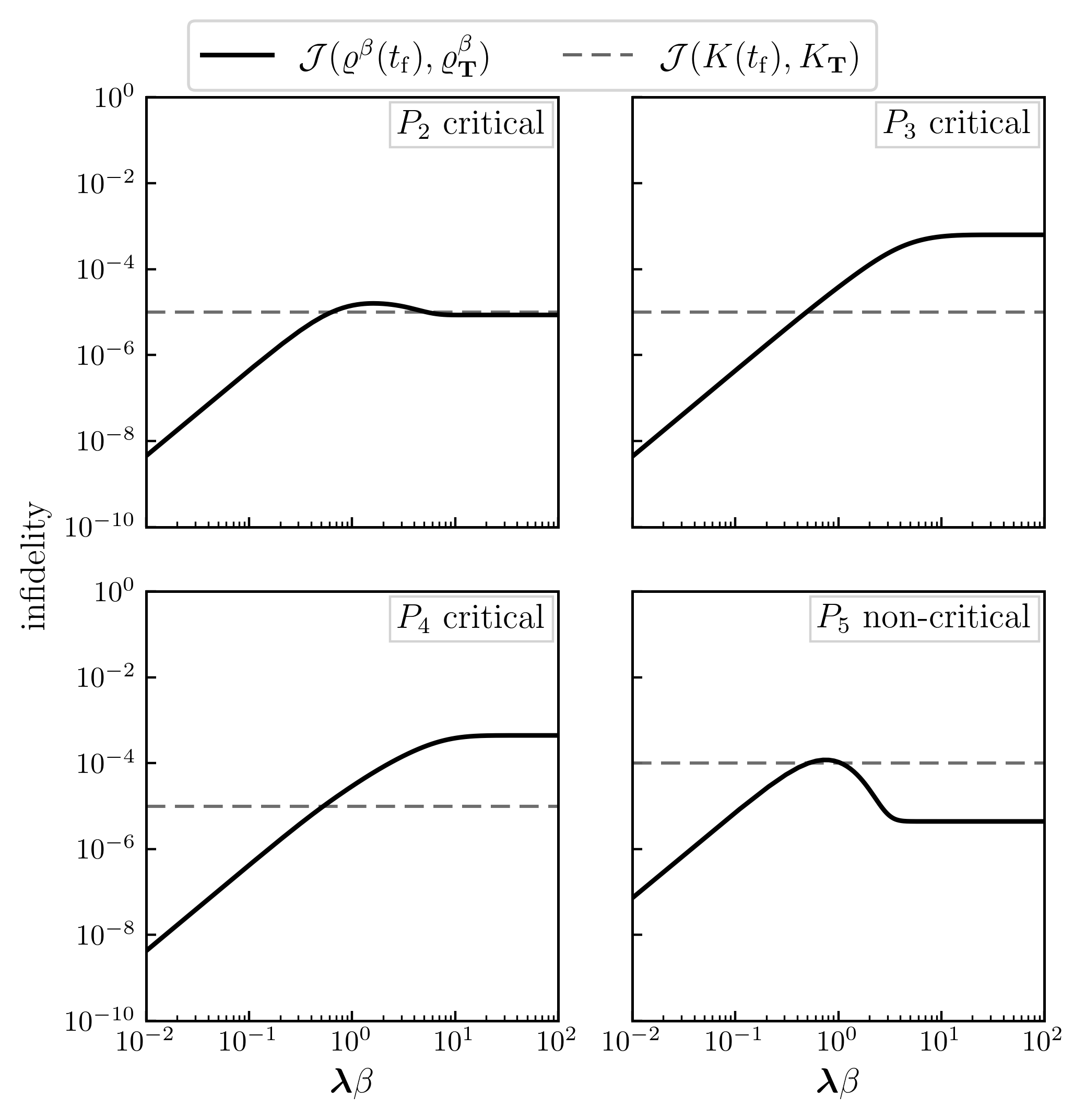}
\caption{Thermal state infidelity (Eq.~\eqref{eq:thermalinfidelity}) as a function of $\boldsymbol{\lambda}\beta$, with the scale $\boldsymbol{\lambda}$ given in Eq.~\eqref{eq:scale}, for the points $P_2,\dots,P_5$ depicted in the phase diagram in Fig.~\ref{fig:labelledtriangle} for a system of $n=11$ spins.
The optimal control solution for the thermal state preparation is obtained by optimizing for the operator infidelity, and the final value obtained in the numerical optimization is depicted as a dashed line.\\
The observations are largely the same as the exemplary points (Fig.~\ref{fig:thermalstates}): the actual figure of merit (the thermal state infidelity ${\cal J}(\varrho^\beta(t_{\mathrm{f}}),\varrho_\mathbf{T}^\beta)$) remains small even though the proxy (the operator infidelity $\mathcal{J}(K(t_{\mathrm{f}}),\Htarget)$) was optimized. For critical points $P_3,P_4$, the thermal state infidelity in the low temperature regime can exceed the operator infidelity, however given a small operator infidelity solution, the worst-case thermal state infidelity remains sufficiently accurate.}
\label{fig:thermalstatessupp}
\end{figure}

\begin{figure}[t!]
\centering
\includegraphics[width=\linewidth]{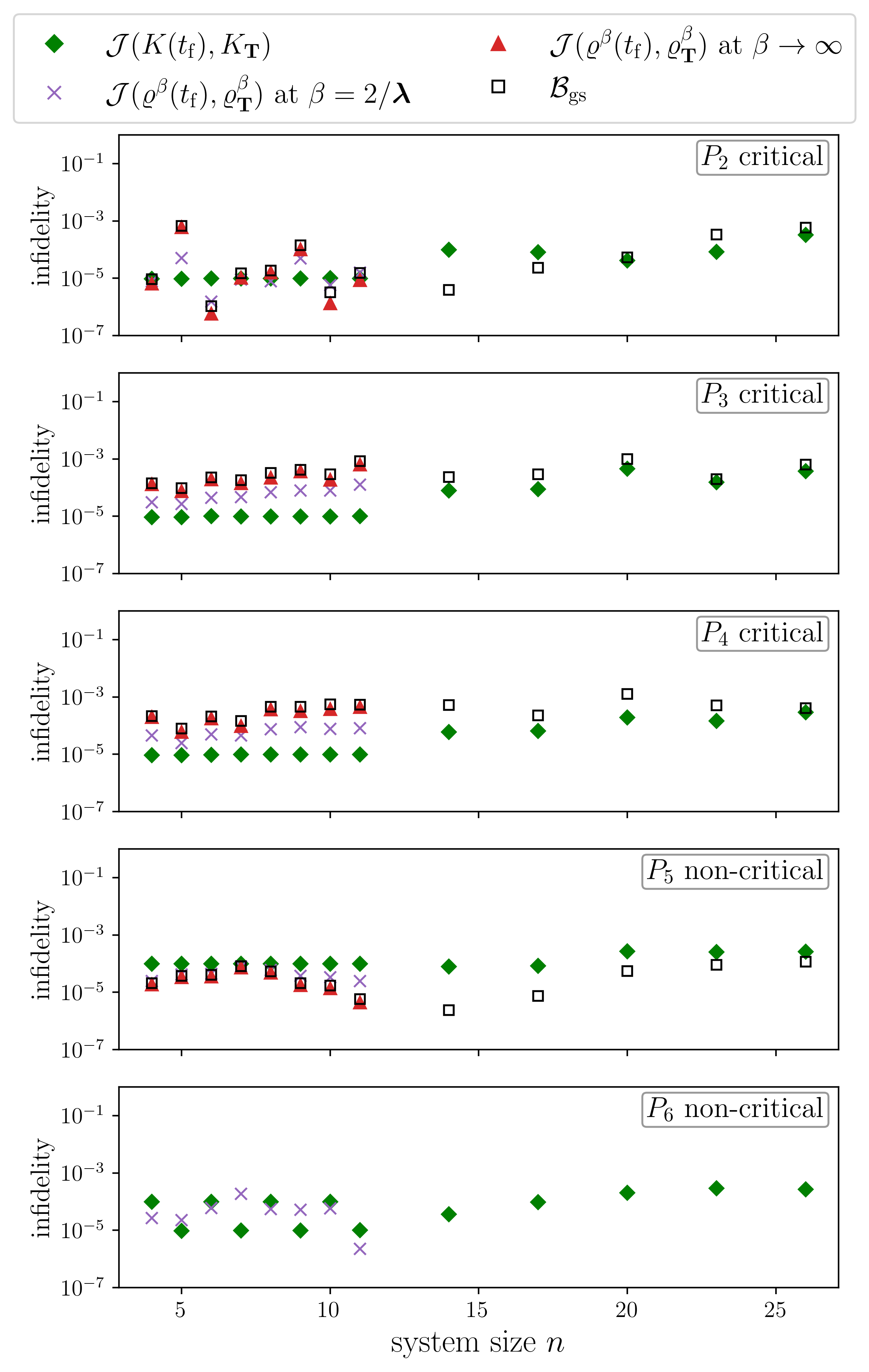}
\caption{Operator infidelity resultant from the optimization for the points $P_2,\dots,P_6$ depicted in Fig.~\ref{fig:labelledtriangle} as a function of system size (green diamonds).
State fidelities obtained with the corresponding optimal control solutions are depicted for system sizes up to $n=11$ spins both for the ground state ($\beta\to\infty$, red triangles) and for the finite temperature $\beta=2/\bm{\lambda}$ (purple crosses) with the scale $\bm{\lambda}$ (Eq.~\eqref{eq:scale}) of $K_\mathbf{T}$. The ground state infidelity upper bound $\mathcal{B}_\mathrm{gs}$ (Eq.~\eqref{eq:gsBound}) is depicted by empty square points. Limited data is shown for non-critical point $P_6$, justified in App.~\ref{app:suppSol}. Similar observations are made to Fig.~\ref{fig:opandth}: in all cases, the operator infidelity is a sufficiently good proxy for the state infidelity to be used as control target.
}
\label{fig:suppopandth}
\end{figure}

\begin{figure}[t!]
\centering
\includegraphics[width=\linewidth]{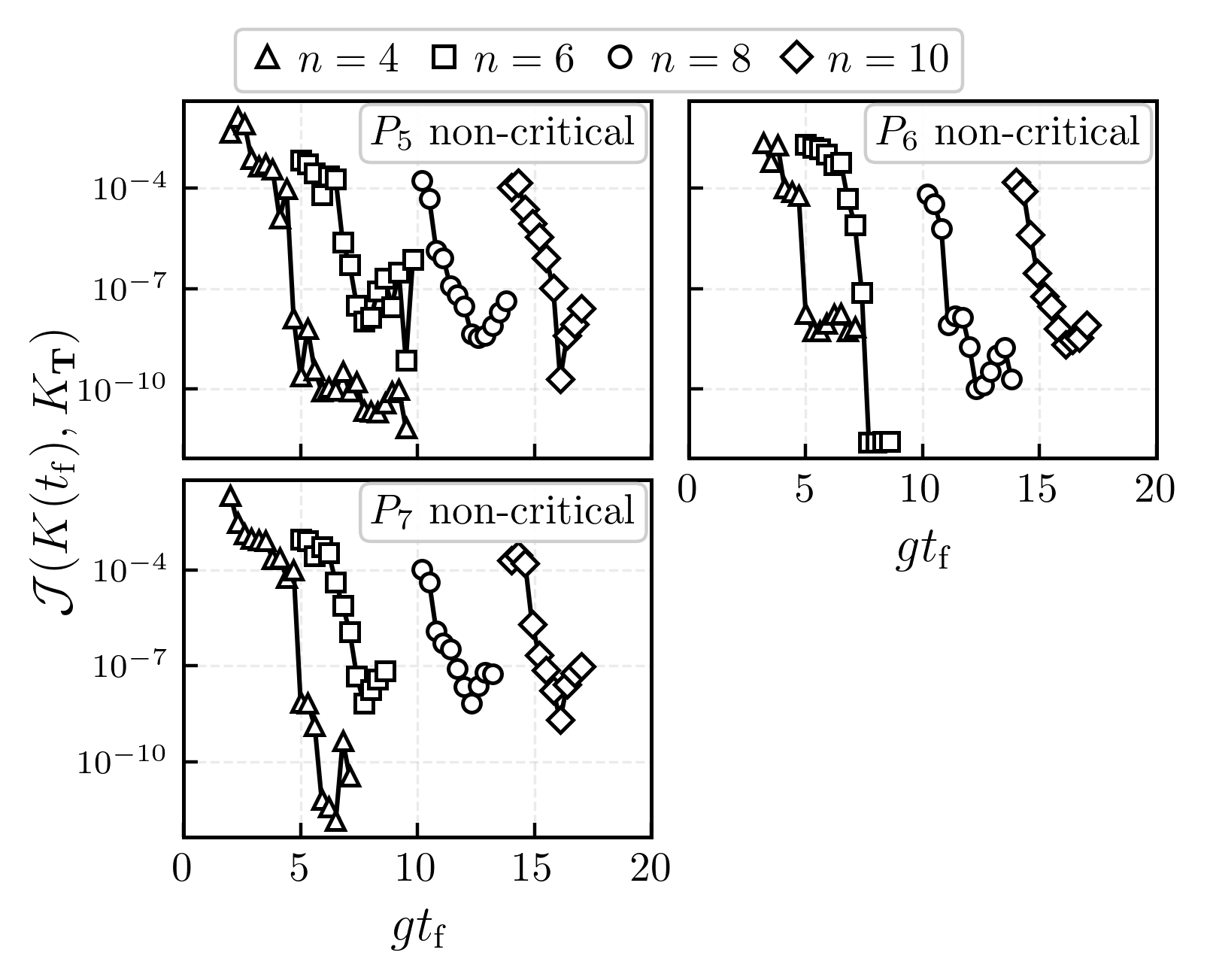}
\caption{Best operator infidelity obtained as result of the optimization as a function of the evolution time $g \tf$, where $g$ is the interaction constant in the system Hamiltonian $H(t)$ (Eq.~\eqref{eq:explicitcontrolham}), for the non-critical points $P_5,P_6,P_7$.
The location of the well-defined drop in the infidelity, which numerically identifies the quantum speed limit, is in close agreement with the data given Fig.~\ref{fig:speedlimits}.}
\label{fig:speedlimitssupp}
\end{figure}

In this section, we show the control solutions for the remaining points of the phase diagram, $P_2\dots,P_6$. As labelled in Fig.~\ref{fig:labelledtriangle}, points $P_2,P_3,P_4$ are critical, and points $P_5,P_6$ are non-critical.

While the non-critical points typically have a sizeable ground state energy gap that does not close with system size, the target $\Htarget$ corresponding to parameters specified by the non-critical point $P_6$ has a notable spectrum. In particular, for even $n$ system sizes, the gap closes to numerical accuracy even at moderate system sizes. For example, with $n=14$ and the scale $\bm{\lambda}=1$ (Eq.~\eqref{eq:scale}), the gap is $\mathcal{O}(10^{-13})$. At intermediate temperature, the ground state is a maximal mixture of the ground and first excited state to a very good approximation. As such, the thermal state fidelity can still be very accurate. However, because of the size of the gap, it is not computationally feasible to thermally realize the actual ground state and compute the infidelity for the limit $\beta\to \infty$.

For odd system sizes $n$, the gap is thermally realizable; however, it remains small and constant for all odd $n$ -- approximately an order of magnitude below the gap for critical targets ({\it i.e.,} points $P_1,\dots,P_4$ in Fig.~\ref{fig:labelledtriangle}) at large system sizes, such as $n=26$. As a result, the numerical deviations of $K(\tf)$ from the true target $\Htarget$ must be sufficiently small, so that the perturbation does not induce substantial mixing among the thermally populated low-lying eigenstates of $\Htarget$.

To that end, we note that the small gaps for the point $P_6$ lack physical relevance -- they arise from a small perturbation from the exactly two-fold degenerate pure $X_jX_{j+1}$ Hamiltonian rather than any form of criticality. 
Because the only requirement is finding a more accurate optimal control solution, rather than increasing the physical evolution $\tf$, this is a matter of optimization-time complexity, and thus the present method can still be used to prepare a target thermal state in such regions, provided the gap is large enough that the ground state can still be distinguished numerically.

Nevertheless, because of the unusually small gaps, we only consider the intermediate temperature state infidelity for this point. This point is thus omitted from Fig.~\ref{fig:thermalstatessupp} (the supplementary plot of Fig.~\ref{fig:thermalstates}), and limited data is shown in Fig.~\ref{fig:suppopandth} (the supplementary plot of Fig.~\ref{fig:opandth}). As with points $P_1$ and $P_7$ in the main text, the operator infidelity $\mathcal{J}(K(\tf),K_\mathbf{T})$ is evidently a good proxy, ensuring that the thermal state infidelity $\mathcal{J}(\varrho^\beta(\tf),\varrho^\beta_\mathbf{T})$ is also small, and the plots show qualitatively similar behaviour for points in critical and non-critical regions, with $P_6$ as an extreme exception.

Fig.~\ref{fig:speedlimitssupp} is the supplementary plot of Fig.~\ref{fig:speedlimits} and shows the drops in operator infidelity required to resolve the quantum speed limit duration for the remaining non-critical points, such that the durations can be read-off the plot. Strikingly, the specific speed limit durations $t_{\text{min}}$ appear to coincide with that of critical points, and additionally grow linearly with system size (as shown in Fig.~\ref{fig:tqslvsN}).

\section{Details of numerical implementation} \label{app:suppOpt}
Here, we provide additional details for the technicalities of the optimization including the equation of motion required for efficient numerical propagation using the von Neumann equation (Eq.~\eqref{eq:vN}) discussed in App. \ref{app:eom}, and analytical gradients derived in App. \ref{app:gradients}. Such details are largely based on the standard implicit control framework \cite{orozco2024quantum}. 
Finally, we end this section with some more subtle technical details including further details of the objective function and discretization of the system Hamiltonian's time-dependence in App. \ref{app:furthertech}.

\subsection{Equation of motion} \label{app:eom}

The control terms,
\begin{equation}
    \left\{Z_i,X_1,X_n,X_jX_{j+1}\right\},
\end{equation}
where $i\in\left\{1,\dots ,n\right\}$ and $j\in\left\{1,\dots ,n-1\right\}$, form the polynomially scaling Lie algebra $\mathfrak{b}$, with elements $\left\{b_j\right\}$ satisfying closed commutation relations (Eq.~\eqref{eq:Lie}). 

Writing the system Hamiltonian as the expansion,
\begin{equation}
    H(t)=\sum_jh_j(t)b_j,
\end{equation}
and,
\begin{equation}
    K(t)=\sum_ja_j(t)b_j,
\end{equation}
the von Neumann equation can be re-written as
\begin{align}
    \dot{K}(t)&=i\left[ K(t),H(t)\right] \\
    &=i\sum_{jk}a_j(t)h_k(t) \left[b_j,b_k\right] \\
    &=\sum_{jkl}a_j(t)h_k(t)\lambda _{lj}^kb_l.
\end{align}
Comparing coefficients for $b_l$ (because of linear independence), gives 
\begin{equation}
    \dot{a}_l(t)=\sum_{jk}\lambda_{lj}^k a_j(t)h_k(t),
\end{equation}
which can be written compactly as
\begin{equation}
    \dot{\mathbf{a}}(t)=G(t)\mathbf{a}(t) ,
    \label{eq:eqofmotion}
\end{equation}
where $\mathbf{a}(t)$ is a time-dependent coefficient vector with elements $\left\{a_j\left(t\right)\right\}$ with length equal to the number of elements in the Lie algebra, and $G(t)$ is a matrix with elements,
\begin{equation}
    G_{lj}(t)=\sum_kh_k(t)\lambda_{lj}^k.
\end{equation}
To that end, the differential equation (Eq.~\eqref{eq:eqofmotion}) provides an efficient numerical representation of the von Neumann equation, owing to the quadratic scaling of elements in $\mathfrak{b}$. Additionally, because $G(t)$ is sparse, Krylov methods can be employed for further computational efficiency.

\subsection{Analytical gradient of the objective function} \label{app:gradients}
As is common practice with GRAPE-inspired optimization, it is often the case that analytical gradients enable more efficient convergence than numerical gradient estimators such as finite differences. 
By defining a target coefficient vector $\mathbf{a}_\mathbf{T}$ with elements $\left\{a_{\mathbf{T},j}\right\}$ such that the control target can be written using the expansion,
\begin{equation}
    K_\mathbf{T}=\sum _ja_{\mathbf{T},j}b_j,
\end{equation}
the operator infidelity $\mathcal{J}(K(\tf),K_\mathbf{T})$ (Eq.~\eqref{eq:objfunction}) takes the form,
\begin{equation}
    \mathcal{J}=1-\frac{\mathbf{a}(\tf)\cdot\mathbf{a}_\mathbf{T}}{\norm{\mathbf{a}(0)}\norm{\mathbf{a}_\mathbf{T}}},
    \label{eq:objfunctionvectors}
\end{equation}
where we have used that $\norm{\mathbf{a}(\tf)}=\norm{\mathbf{a}(0)}$. Eq.~\eqref{eq:objfunctionvectors} which can be efficiently evaluated.

We provide two derivations: the analytical gradient of $\mathcal{J}$ with respect to the discretized time-dependence of the system Hamiltonian, $\left\{h_j\left(t\right)\right\}$ (Sec.~\ref{app:wrtHam}), and with respect to the coefficients $\left\{c_j\right\}$ that specify the initial condition Eq.~\eqref{eq:initialcond} (Sec.~\ref{app:wrtZinit}).

\subsubsection{With respect to system Hamiltonian}\label{app:wrtHam}

Let $h_{k,m}$ be constant within a time slice $m$, representing a discretized time-dependence $h_k(t)$. In this time-slice, the matrix $G(t)$ is given by $G_m=\sum_kh_{k,m}\Lambda_k$ where $\Lambda_k$ is a $d\times d$ matrix built from structure coefficients $\lambda_{lj}^k$, and $d$ is the number of elements in the Lie algebra $\mathfrak{b}$. With a total of $M$ time-slices, where each time-slice duration is $\tau_m=t_{m}-t_{m-1}$, the unitary propagator from the solution of Eq.~\eqref{eq:eqofmotion}, is given by $\mathcal{U}_m=e^{G_m\tau_m}$. One can write,
\begin{align} 
 \frac{\partial \mathcal{J}}{\partial h_{k,m}} &= -\frac{\partial }{\partial h_{k,m}}\left(\frac{\mathbf{a}_\mathbf{T}\cdot\mathbf{a}(\tf)}{\norm{\mathbf{a}(0)}\norm{\mathbf{a}_\mathbf{T}}}\right) \\ 
 &= -\frac{\partial }{\partial h_{k,m}}\left(\frac{\mathbf{a}_\mathbf{T}^\dagger\mathbf{a}(\tf)}{\norm{\mathbf{a}(0)}\norm{\mathbf{a}_\mathbf{T}}}\right) \\
 &=-\frac{\partial }{\partial h_{k,m}}\left(\frac{\mathbf{a}_\mathbf{T}^\dagger \mathcal{U}_M\mathcal{U}_{M-1}\dots \mathcal{U}_1\mathbf{a}(0)}{\norm{\mathbf{a}(0)}\norm{\mathbf{a}_\mathbf{T}}}\right) \\
 &=-\frac{1}{\norm{\mathbf{a}(0)}\norm{\mathbf{a}_\mathbf{T}}}\frac{\partial }{\partial h_{k,m}}\left(\mathbf{a}_\mathbf{T}^\dagger \mathcal{U}_M\mathcal{U}_{M-1}\dots \mathcal{U}_1\mathbf{a}(0)\right) \\
 &=-\frac{1}{\norm{\mathbf{a}(0)}\norm{\mathbf{a}_\mathbf{T}}}\mathbf{b}^{\dagger }_{m+1}\frac{\partial \mathcal{U}_m}{\partial h_{k,m}}\mathbf{a}_{m-1}
\end{align}
where $\mathbf{b}^{\dagger }_{m+1}=\mathbf{a}_\mathbf{T}^{\dagger }\mathcal{U}_M\mathcal{U}_{M-1}\dots \mathcal{U}_{m+1}$ (backwards propagation) and $\mathbf{a}_{m-1}=\mathcal{U}_{m-1}\dots \mathcal{U}_2\mathcal{U}_1\mathbf{a}\left(0\right)$ (forwards propagation).

For any differentiable matrix-valued function $X(\alpha)$,
\begin{equation}
    \frac{\partial }{\partial \alpha }e^{X\left(\alpha \right)}=\int _0^1e^{\left(1-s\right)X\left(\alpha \right)}\frac{\partial X}{\partial \alpha }e^{sX\left(\alpha \right)}ds.
\end{equation}
For $\partial \mathcal{U}_m/\partial h_{k,m}$, one has $\mathcal{U}_m=e^{G_m\tau_m}$ and $\alpha=h_{k,m}$. Thus $\partial X/\partial \alpha$ becomes $\tau_m \partial G_m/\partial h_{k,m}=\tau_m\Lambda_k$, and one can write
\begin{align} 
 \frac{\partial \mathcal{U}_m}{\partial h_{k,m}} &= \tau _m\int _0^1e^{\left(1-s\right)G_m\tau _m}\Lambda _ke^{sG_m\tau _m}ds \\ 
 &= \mathcal{U}_m\tau _m\int _0^1e^{-sG_m\tau _m}\Lambda _ke^{sG_m\tau _m}ds \\
 &=\mathcal{U}_m\int _0^{\tau _m}e^{-tG_m}\Lambda _ke^{tG_m}dt,
\end{align}
where in the last line the substitution $s=t/\tau_n$ is made. Taylor expanding the exponentials and computing the integration gives, 
\begin{equation}
    \frac{\partial \mathcal{U}_m}{\partial h_{k,m}}=\left(\tau _m\Lambda _k\mathcal{U}_m-\frac{\tau _m^2}{2}\left[G_m,\Lambda _k\right]\mathcal{U}_m\right)+\mathcal{O}\left(\tau _m^3\right),
\end{equation}
and finally,
\begin{equation}
    \frac{\partial \mathcal{J}}{\partial h_{k,m}} \approx-\frac{1}{\norm{\mathbf{a}(0)}\norm{\mathbf{a}_\mathbf{T}}}\mathbf{b}^{\dagger }_{m+1}(\tau _m\Lambda _k-\frac{\tau _m^2}{2}\left[G_m,\Lambda _k\right])\mathbf{a}_{m}
\end{equation}

\subsubsection{With respect to initial condition}\label{app:wrtZinit}

The coefficients $\left\{c_i\right\}$ that specify the initial condition,
\begin{equation}
    K_0=\sum_{j=1}^{n} c_jZ_j + c_{n+1}\mathbf{Z},
\end{equation}
are additionally found during the optimization. Note that $c_{n+1}$ was given as $c_0$ in the main text (Eq.~\eqref{eq:Zinit}).  Let $A=\norm{\mathbf{a}_\mathbf{T}}$, $V=\norm{\mathbf{a}(\tf)}$ and $F=\mathbf{a}_\mathbf{T}\cdot\mathbf{a}(\tf)$, such that $\mathcal{J}=1-F/AV$. For each $\ell$ component in $\mathbf{a}(\tf)$,
\begin{align}
    \frac{\partial\mathcal{J}}{\partial a_\ell(\tf)}&=-\frac{1}{A}\frac{\partial }{\partial a_{\ell }\left(\tf\right)}\left(\frac{F}{V}\right)\\
    &=-\frac{1}{A}\left(\frac{V\partial _{\ell }F-F\partial _{\ell }V}{V^2}\right),
\end{align}
where $\partial _{\ell }=\partial / \partial a_{\ell }(\tf)$. With $F=\sum _ja_{\mathbf{T},j}a_j\left(\tf\right)$, one has $\partial _{\ell }F=a_{\mathbf{T},\ell }$.
Additionally with $ V=\left[\sum _ja_j^2\left(\tf\right)\right]^{\frac{1}{2}}$, one has
\begin{equation}
    \partial _{\ell }V=\frac{1}{2}\left[\sum _ja_j^2\left(\tf\right)\right]^{-\frac{1}{2}}2a_{\ell }\left(\tf\right) =\frac{a_{\ell }\left(\tf\right)}{V} .
\end{equation}
Therefore, we have
\begin{align}
    \frac{\partial\mathcal{J}}{\partial a_\ell(\tf)}&=-\frac{1}{A}\frac{Va_{\mathbf{T},\ell }-F[a_{\ell }\left(\tf\right)/V]}{V^2}\\
    &=-\frac{a_{\mathbf{T},\ell }}{AV}+\frac{Fa_{\ell }\left(\tf\right)}{AV^3},
\end{align}
or in column vector form,
\begin{equation}
    \nabla_{\mathbf{a}(\tf)}\mathcal{J}=-\frac{\mathbf{a}_\mathbf{T}}{AV}+\frac{F}{AV^3}\mathbf{a}(\tf)=\mathbf{v}_\mathbf{T}.
\end{equation}
Let $\left\{i_1,\dots ,i_{n+1}\right\}\subset \left\{1,\dots ,d\right\}$ be the index set that belongs to $\mathfrak{h}$, the maximal abelian set in $\mathfrak{m}$, where  $d$ is the number of elements in the Lie algebra $\mathfrak{b}$. $\mathfrak{h}$ contains $n+1$ elements. Let $P$ be a selection matrix with elements $P_{\ell,i}=\delta_{\ell,h_i}$, such that 
\begin{equation}
    \mathbf{a}(0)=P\mathbf{c},
\end{equation}
where $\mathbf{c}=\left(c_1,\dots ,c_{n+1}\right)^{\top }$ is a vector of coefficients for the initial condition. Therefore, $\partial \mathbf{a}(0)/\partial\mathbf{c}=P$,
\begin{align}
    \frac{\partial\mathcal{J}}{\partial\mathbf{c}}&=\frac{\partial\mathcal{J}}{\partial\mathbf{a}(\tf)}\frac{\partial\mathbf{a}(\tf)}{\partial\mathbf{c}}\\
    &=\frac{\partial\mathcal{J}}{\partial\mathbf{a}(\tf)}\mathcal{U}_M\mathcal{U}_{M-1}\dots \mathcal{U}_1\frac{\partial\mathbf{a}(0)}{\partial\mathbf{c}}. 
\end{align}
Noting that $\partial\mathcal{J/\partial\mathbf{a}}(\tf)$ is the Jacobian row corresponding to the column gradient $\nabla_{\mathbf{a}(\tf)}\mathcal{J}$, {\it i.e.,}  $\partial\mathcal{J/\partial\mathbf{a}}(\tf)=(\nabla_{\mathbf{a}(\tf)}\mathcal{J})^\top$, we can write
\begin{align}
    \frac{\partial\mathcal{J}}{\partial\mathbf{c}}&=\frac{\partial\mathcal{J}}{\partial\mathbf{a}(\tf)}\mathcal{U}_M\mathcal{U}_{M-1}\dots \mathcal{U}_1\frac{\partial\mathbf{a}(0)}{\partial\mathbf{c}}\\
    &=\mathbf{v}_\mathbf{T}^\top \mathcal{U}_M\mathcal{U}_{M-1}\dots \mathcal{U}_1P \\
    &=\mathbf{v}_0^\top P,
\end{align}
where $\mathbf{v}_0^\top=\mathbf{v}_\mathbf{T}^\top \mathcal{U}_M\mathcal{U}_{M-1}\dots \mathcal{U}_1$.

\subsection{Further technicalities}\label{app:furthertech} 
\subsubsection{Objective function}
The operator infidelity $\mathcal{J}(K(\tf),K_\mathbf{T})$ (Eq.~\eqref{eq:objfunction}) is minimal if and only if the operators $K(\tf)$ and $K_\mathbf{T}$ coincide up to a positive scalar factor, {\it i.e.} $K(\tf)=cK_\mathbf{T}$.
To achieve the actual goal, where $K(\tf)$ and $K_\mathbf{T}$ exactly coincide for $\mathcal{J}(K(\tf),K_\mathbf{T})=0$, one can rescale the initial condition $K_0$. With, 
\begin{align}
    f(c)&=\norm{K(\tf)-cK_\mathbf{T}}^2 \\
    &=\norm{K(\tf)}^2-2c\left\langle K\left(\tf\right),K_\mathbf{T}\right\rangle +c^2\norm{K_\mathbf{T}}^2 ,
\end{align}
Given Hermitian operators, the expression is minimized by scalar
\begin{equation}
    c=\frac{\text{tr}(K(\tf)K_\mathbf{T})}{\text{tr}(K_\mathbf{T}^2)} ,
\end{equation}
satisfying $f'(c)=0$. With this expression being efficiently evaluable using the coefficient vectors, Eq.~\eqref{eq:Isol} can be restated as
\begin{equation}
    U(\tf)\ c^{-1}K_0\ U^\dagger(\tf)=c^{-1}K(\tf)\approx K_\mathbf{T}
\end{equation}
enabling the required unitary mapping.

\subsubsection{Discretization}

There are points in the phase diagram (Fig.~\ref{fig:labelledtriangle}) that are more difficult to find desired optimal control solutions than others. These are typically the critical points, $P_1,\dots,P_4$. Because, regardless of the point in the phase diagram, there is a guaranteed solution that maps the initial thermal state $\varrho^\beta(0)$ to the target thermal state $\varrho_\mathbf{T}^\beta$ via a unitary generated by the system Hamiltonian (Eq.~\eqref{eq:explicitcontrolham}) (owing to Eq.~\eqref{eq:reach}), the difficulty can be attributed to a challenging optimization landscape rather than a fundamental barrier to reaching the solution. 

To address such points, there are some helpful aspects to consider. Firstly, one can select a finer discretization of the time-dependent factors $\left\{h_j\left(t\right)\right\}$. 
At the cost of longer computational running time, the optimization routine seems to escape unwanted local minima more effectively. 
This is particularly useful for resolving the quantum speed limit, Figs.~\ref{fig:speedlimits} and \ref{fig:speedlimitssupp}, in which optimal control solutions with operator infidelities of $\mathcal{O}(10^{-10})$ are typically required. 
In this data, a discretization of $N=150n$ was found to be sufficient, where optimal control solutions of sufficient accuracy to resolve the quantum speed limit are often found with a single run of the algorithm. This is the main reason why the data is presented only up to moderate system sizes (up to $n=10$), where the computational running times at this discretization are reasonable.

Secondly, the optimization routine can be run for a less optimal discretization, and then one can improve upon the yielded solution. In particular, with a discretization of $N=20n$, the optimization will almost certainly converge to an infidelity that is considerably higher than what is desired, especially for large system sizes $n\ge 20$. Taking this suboptimal solution as the initial guess for the next optimization, with the same discretization, one can iteratively improve on the original solution. Alternatively, one can start with the optimal control solution for a nearby point (in the phase diagram), which should have an easier optimization landscape, and then perturbatively progress towards the desired point. Namely, one starts with the initial guess as the solution of the previous optimization and specifies the target as the perturbed point.

In all cases, the only restriction to finding a better solution is time. One can always improve upon the yielded solution if time allows, as opposed to restrictions involving memory (as is the case for explicit state-vector or density matrix control methods). To that end, a solution must always be accepted within a reasonable time frame. For the data presented in this paper, Figs.~\ref{fig:opandth} and \ref{fig:suppopandth}, operator infidelities of $\mathcal{O}(10^{-4})$ (resulting in worst-case critical region thermal state infidelities between $\mathcal{O}(10^{-3})$ and $\mathcal{O}(10^{-2})$) are typically accepted solutions given the accuracy of today's quantum hardware.

\vspace{1cm}
\bibliography{library_updated.bib}
\end{document}